\documentstyle[pra,aps]{revtex}
\twocolumn

\begin{document}
\draft
\title{Quantum Kinetic Theory III: Simulation of the Quantum
Boltzmann
Master Equation}
\author{D.~Jaksch$^{1}$, C.W.~Gardiner$^{2}$ and P.~Zoller$^{1}$}
\address{$^1$ Institut f{\"u}r Theoretische Physik,
Universit{\"a}t Innsbruck, 6020 Innsbruck, Austria}
\address{$^2$ Physics Department, Victoria University, Wellington,
New Zealand}

\maketitle

\tighten
\narrowtext

\begin{abstract}
We present results of simulations of a {\em quantum Boltzmann master
equation} (QBME) describing the kinetics of a dilute Bose gas
confined
in a trapping potential in the regime of Bose condensation.  The QBME
is the simplest version of a quantum kinetic master equations derived
in previous work.  We consider two cases of trapping potentials: a 3D
square well potential with periodic boundary conditions, and an
isotropic harmonic oscillator. We discuss the stationary solutions
and
relaxation to equilibrium.  In particular, we calculate particle
distribution functions, fluctuations in the occupation numbers, the
time between collisions, and the mean occupation numbers of the
one-particle states in the regime of onset of Bose condensation.
\end{abstract}

\section{Introduction}

In the previous two papers \cite{QKME1,QKME2}, which we will refer to
as QKI and QKII, a fully quantum mechanical kinetic theory for Bose
gases was developed. One of the simplest versions of the quantum
kinetic master equation (QKME) neglects all spatial dependence, and
yields a master equation, which we have named the quantum Boltzmann
master equation (QBME).  In contrast to the familiar quantum
Boltzmann
equation (QBE) \cite{BECKagan,BECSemikoz}, which is an equation of
the
single particle distribution function, the QBME is an $N$--atom
stochastic equation.  The aim of the present paper is to present
results of numerical simulations of this equation for finite size
systems consisting typically of a few hundred atoms. Although the
exclusion of the spatial dependence is an extreme simplification,
these simulations will give us a first orientation about the kind of
solutions the QKME will yield. These simulations can thus serve as a
guideline for the type of approximations of the QKME one may use to
find numerical solutions of this much more interesting, but unwieldy,
equation.

Furthermore, we will concentrate our attention on those results of
the
QBME which cannot be obtained using equations like the quantum
Boltzmann equation (QBE) .  We also restrict our work to the region
of
temperatures which are less than or not much higher than the critical
temperature of the gas, because at much higher temperatures quantum
effects do not play a crucial role and simulations of the classical
Boltzmann equation, which is valid in that case, have already
been performed \cite{TraSimMurray}.

The QBME is a genuine $N$--atom equation like the QKME, but it
neglects
all the coherences contained in the QKME---it is thus intermediate
between the QKME and the description of the system with kinetic
equations for single-particle distribution functions.  Its
irreversibility comes from the Markov assumption employed in deriving
the QKME.

The paper is organized as follows. In Sec.~\ref{QBMEder} we review
the
derivation of the QBME in QKI \cite{QKME1}, discuss properties of the
QBME, and compare it with the QBE.  Furthermore, we give a
brief description of the simulation algorithm.  In Secs.~\ref{3Dbox}
and \ref{HO} we apply the QBME to study a Bose gas confined in a 3D
box and
in a 3D harmonic oscillator.  In particular, we discuss simulations
results for
thermodynamic quantities, the mean time between collisions, and the
fluctuations of the occupation numbers of the condensate.  For the 3D
harmonic oscillator we also simulate a gas that is evaporatively
cooled.

\section{The quantum Boltzmann master Equation}
\label{QBMEder}

In this section we will first summarize the derivation of the quantum
Boltzmann master equation as given in QKI \cite{QKME1}. Furthermore,
we discuss properties of this equation and its solutions which are
relevant for our numerical studies presented in Secs.~\ref{3Dbox} and
\ref{HO}, and conclude with a comparison of the QBME with the QBE.

\subsection{Derivation and validity of the QBME}

The second quantized form of the Hamilton operator for a Bose gas 
with pair particle interaction can be written $H=H_0 + H_I$, where
\begin{equation} H_{0}=\sum_{{{\bf {m}}}_{i}} \hbar
\omega_{{{\bf {m}}}_{i}} a_{{{\bf {m}}}_{i}}^{\dagger} a_{{{\bf
{m}}}_{i}} ,\end{equation} \begin{equation}
H_{I}=\frac{1}{2} \sum_{{{\bf {m}}}_{1},{{\bf {m}}}_{2},{{\bf
{m}}}_{3},{{\bf {m}}}_{4}}
U_{{{\bf {m}}}_{1},{{\bf {m}}}_{2},{{\bf {m}}}_{3},{{\bf {m}}}_{4}}
a^{\dagger}_{{{\bf {m}}}_{1}} a^{\dagger}_{{{\bf {m}}}_{2}} a_{{{\bf
{m}}}_{3}}
a_{{{\bf {m}}}_{4}} .\end{equation} Here $H_{0}$ is the system
Hamilton operator of
the non interacting Bose gas where $a_{{{\bf {m}}}_{i}}^{\dagger}$ is
the
creation operator of a particle in the eigenstate of $H_0$ labeled
${{\bf {m}}}_{i}$ with energy $\hbar \omega_{{{\bf {m}}}_{i}}$.  The
trapping
potential is included in $H_0$.

The interaction Hamiltonian $H_{I}$ describes two body interactions
in
the Bose gas. In the regime we want to study, only $s$-wave
scattering
plays an important role, allowing us to write
\begin{eqnarray} \label{pot}
&& 	U_{{{\bf {m}}}_{1},{{\bf {m}}}_{2},{{\bf {m}}}_{3},{{\bf
{m}}}_{4}}
\nonumber\\
&& \qquad
=\frac{4 \pi\hbar^2 a}{m}
\int\limits_{R^3} d^3\!x \: \Psi^{*}
_{{{\bf {m}}}_{1}}({{\bf {x}}})
\Psi^{*}_{{{\bf {m}}}_{2}}({{\bf {x}}}) \Psi_{{{\bf {m}}}_{3}}({{\bf
{x}}})
\Psi_{{{\bf {m}}}_{4}}({{\bf {x}}}).
\end{eqnarray}
In Eq.~(\ref{pot}) $\Psi_{{{\bf {m}}}_{i}}({{\bf {x}}})$ denotes an
eigenfunction of the trapping potential in coordinate space labeled
by quantum numbers $ i$.  Below we will specify the potential to be a
3D box with periodic boundary conditions (Sec.~\ref{3Dbox}), or a 3D
isotropic oscillator (Sec.~\ref{HO}), and will give expressions for
the matrix elements $ U_{{{\bf {m}}}_{1},{{\bf {m}}}_{2},{{\bf
{m}}}_{3},{{\bf {m}}}_{4}}$
for these specific cases.  For convenience we will use the
notation $i$ instead of ${{\bf {m}}}_{i}$ below. The scattering
length of the gas
is $a$ and the mass of the gas particles is $ m$. We will treat
systems with finite number of particles $N$.  This is the starting
point from which the QKME is derived in QKI. The following
assumptions
and approximations are made:

\subsubsection{The forward scattering terms}

All the terms of $H_{I}$ (see QKI Eq.~(67)) which
commute with the system Hamiltonian $H_{0}$ describe forward
scattering and give rise to the mean field. These terms can be
included with $H_{0}$.  Forward scattering does not change the
occupation of the one-particle eigenstates, so we will neglect the
influence of these terms on the eigenstates of $H$ in the
simulations.

\subsubsection{The collision terms}

The remaining terms in $H_I$ describe collisions which change the
occupation numbers of the one-particle states of the trapping
potential.  We assume that this part of $H_{I}$ can be treated
perturbatively, using the Born approximation and the Markov
approximation (QKI Sec.~IV C 3).  The Born approximation is valid
when
the interaction between the particles is small compared to the system
Hamiltonian $H_0$ \cite{Tmatrix}.  For the Markov approximation to be
valid it is required that the frequency spectrum is effectively
continuous which means that the separation between the energy levels
is much smaller than the energy range of occupied states.  The use of
the Markov approximation gives the QKME its irreversible character.
We
will neglect the influence of collisional shifts on eigenstates of
$H$.

\subsubsection{Reduction of the QKME to the QBME}

To reduce the QKME to the QBME it is assumed that the coherent terms
 (i.e.~Hamiltonian terms in QKI Eq.~(77)) can be neglected.  The QBME
 is an equation for the diagonal elements $w_{{{\bf {n}}}} \equiv
 {{\left\langle {{{{\bf {n}}}}} \right|} {\rho} {\left| {{{{\bf
{n}}}}} \right\rangle}}$ of the density operator and takes the
 following form (QKI Eq.~(101)):
\begin{eqnarray}
\dot{w}_{{{\bf {n}}}} & = & -\frac{\pi}{\hbar} \sum_{1234} \delta
\left(\hbar(\omega_{1} + \omega_{2} - \omega_{3} - \omega_{4})\right)
\left| U_{1 2 3 4} \right|^{2} \nonumber \\ &&\times \left\{ n_{1}
n_{2} \left( n_{3}+1 \right) \left( n_{4}+1 \right) \left[ w_{{{\bf
{n}}}}-
w_{{{\bf {n}}}+{{\bf {e}}}_{1234}} \right] \right.  \nonumber \\ & &
+
\left. \left( n_{1} +1\right) \left( n_{2} + 1 \right) n_{3} n_{4}
\left[ w_{{{\bf {n}}}}- w_{{{\bf {n}}}-{{\bf {e}}}_{1234}} \right]
\right\}
.\label{QBME}
\end{eqnarray}
Here ${\left| {{{\bf {n}}}} \right\rangle}={\left| {n_0,n_1,n_2,...}
\right\rangle}$ is a Fock state of the $
N$-particle system, giving the occupation numbers $n_i$ of the
eigenstates $\Psi_i({{\bf {x}}})$ and ${{\bf {n}}}$ denotes the
vector
consisting of the occupation numbers $n_i$.  The vector
${{\bf {e}}}_{1234}$ is defined similarly to ${{\bf {n}}}$ as
\begin{equation}
{{\bf
{e}}}_{1234}=[0,...0,\stackrel{1}{1},0,...0,\stackrel{2}{1},0...0,
\stackrel{3}{ -1},0,...0, \stackrel{4}{-1},0,...0], \end{equation}
which
describes two particle collisions. The state
${\left| {{{\bf {n}}}-{{\bf {e}}}_{1234}} \right\rangle}$ can thus be
reached from ${\left| {{{\bf {n}}}} \right\rangle}$ by
the collision $1+2 \rightarrow 3+4$.

The $\delta$--function in the discrete sum of the QBME (\ref{QBME}) 
has its origin in the use of the Markov approximation as outlined in
QKI.
Since we do not replace these sums by integrals in our simulations
this
$\delta$--function requires interpretation.  We concentrate energy
regions
of $\Delta e$ to one single discrete energy level. The energy
interval is described by the properties of the closest one-particle
state of $H_0$. The choice of $\Delta e$ depends on the trapping
potential and is such that each of the one-particle states serves as
one of the discrete energy levels with energy $e_i$ that determine
the
properties of particles within the energy range $[e_i-\Delta
e/2,e_i+\Delta e/2]$. Implicitly this includes the interpretation of
$n_i$ as being an integral over a smooth distribution function
$f(e)$:
\begin{equation} 
n_i=\int_{e_i-\Delta e/2}^{e_i+\Delta e/2} de \: \frac{f(e)}{\Delta
e}
\label{intn}
\end{equation} where $f(e)$ gives the number of particles occupying a
state with
energy $e$. Among the degenerate one-particle eigenstates the
particles are distributed according to similar arguments as in
Eq.~(\ref{intn}).  The $\delta$--function in the QBME (\ref{QBME})
has, therefore, to be interpreted as \begin{equation}
\delta(e)=\frac{{\delta_{{e},{0}}}}{\Delta e} \end{equation} and
$w_{{{\bf {n}}}}$ in the
QBME (\ref{QBME}) is the probability of finding $n_i$ particles
within
the energy interval $[e_i-\Delta e/2,e_i+\Delta e/2]$.
${\delta_{{x},{y}}}$ denotes the Kronecker delta.

In QKI it is shown that the kernel of the integral where
the Markov approximation is made (QKI Eq.~(68)) has a width given by
the temperature
$\hbar/kT$. This width determines the range of possible outcomes of a

collision. As long as $\hbar/kT$ is much smaller than the time
between two
collisions the free evolution after the kernel has reached zero
will fix the energy of the particles within a range of $\hbar/t_{\rm
coll}$
before the next collision occurs. We already assumed that it is
possible to
describe the system in terms of one-particle eigenstates of the
trapping
potential which is only valid if the level broadening coming from the
collisions is much less than the level spacing. 
Hence, we are able to decide which of our one-particle states 
describes the properties of a particle best before this particle 
collides again.

\subsubsection{Discussion}

Since there are no classical assumptions in deriving the QBME
(\ref{QBME}), it should be valid even when the Bose gas becomes
degenerate, within the limits of the approximations made in its
derivation. The quantum statistics is contained in the $1+n_i$
factors
in Eq.~(\ref{QBME}).  This allows us to study the onset of BEC, in
the
sense of obtaining a macroscopic occupation in the ground state
\cite{senseBEC}, and in particular finite number effects which are
important when the number of atoms is not large.

The QBME (\ref{QBME}) is a full $N$--particle equation in the form of
a
stochastic master equation which describes $N$ particles interacting
with each
other by two particle collisions. These collisions are responsible
for
the equilibration process.  In contrast, the QBE (see
Sec.~\ref{CompQBMEQBE}) considers the motion of one particle
interacting
with a mean distribution of the other atoms in the gas.

No mean field effects are included 
in the present form of the QBME (\ref{QBME}).
As soon as the temperature $T$ of the gas is far below the
critical temperature $T_{c}$ and most of the particles have
accumulated in the 
ground state, the mean field produced by these condensed  particles
must be taken into account. Furthermore, the derivation of the QBME assumes that the width of the energy levels and collisional shift in addition to the mean field
is small relative to the level
spacing $\Delta e$.

\subsection{Quantities of interest}

For comparison with the simulations discussed in the following
sections, we summarize below properties of the stationary solutions, the particle distributions and collision times.

\subsubsection{Stationary solution}

The QBME conserves energy $E$ and number of particles $N$.
According to QKI the stationary solution of the QBME (\ref{QBME})  is

\begin{equation}
w_{{{\bf {n}}}}=\mbox{constant},
\label{statsol}
\end{equation}
corresponding to a microcanonical ensemble.

We will also compare our simulations results with the grand canonical
ensemble.
For the mean occupation numbers one obtains (compare QKI Sec.~V A 2)
\begin{equation}
{\left\langle {n_i}
\right\rangle}=\frac{1}{\exp\-\left(\frac{\hbar\omega_{i}-\mu}{kT}\right)-1}.
\label{BEDstat}
\end{equation}
In this case $T$ is the temperature and $\mu$ is the chemical
potential of
the system in the grand canonical ensemble. 
Given the mean energy of the system $E$ and the mean number of
particles
$N$ we can solve the two equations 
\begin{mathletters}
\begin{eqnarray}
N &=& \sum_i \frac{1}{\exp\-\left(\frac{\hbar\omega_{i}-\mu}{kT}
\right)-1}
\\
E &=& \sum_i \frac{\hbar
\omega_i}{\exp\-\left(\frac{\hbar\omega_{i}-\mu}{kT}\right)-1}
\end{eqnarray}
\label{grca}
\end{mathletters}
for $\mu$ and $T$ numerically. 
We will compare this result below with the one we get from
our simulations. 
In the framework of the QBME these grand canonical results are
obtained 
if we assume that in steady state the expectation values 
of the $n_{i}$ factorize (which is an approximation).

\subsubsection{Particle distributions}
\label{partdis}

The QBME is a stochastic equation for the diagonal elements of the
density
operator in the basis of the eigenstates of $H_{0}$. We are
interested in calculating
the probability distribution of particles in the one-particle states.
They are defined as
\begin{equation}
W_i(j)=\sum_{\stackrel{\scriptsize {{\bf {n}}}}{\scriptsize n_i=j}}
w_{{\bf n}},
\end{equation}
and give the probability of finding $j$ particles in the one-particle
eigenstate labeled $i$. The sum runs over all ${{\bf {n}}}$ with
$\sum_i n_i=N$ and $\hbar \sum_i \omega_i n_i=E$, the constant number
of particles in the gas and the energy of the system, respectively.
We will compute these distributions for the 3D box in
Sec.~\ref{3D:particledistributions}. 

For highly excited states $i$, whose mean occupation number is much
less
than $1$ the probability $W_i(j)$ is only substantially different
from zero for 
$j=0$ and $j=1$, which leads to ${\left\langle {n_i^2}
\right\rangle}-{\left\langle {n_i} \right\rangle}^2 \approx {\left\langle {n_i} \right\rangle}$.
On the other hand, approximate expressions can be derived for low
lying states,
including the ground state, on the following arguments.  Assuming
that there is no restriction on how the particles are distributed
among degenerate energy levels we can write $W_i(j)$ in terms of
energy
levels \cite{Huang}
\begin{equation} 
W_{\bar{i}}(j)=\frac{1}{Z}\sum_{\stackrel{\scriptsize
\bar{n}}{\scriptsize 
g_{\bar{i}} n_{\bar{i}} =j}} \prod_{\bar{l}} \frac{(g_{\bar{l}}
n_{\bar{l}}
+g_{\bar{l}}-1)!}{(g_{\bar{l}}n_{\bar{l}})!(g_{\bar{l}}-1)!}.
\label{flu}
\end{equation} 
Here $\bar{l}$ are sets of indices of degenerate eigenlevels,
$g_{\bar{l}}$ is the number of elements of $\bar{l}$, and
$g_{\bar{l}}n_{\bar{l}}$
gives the total number of particles in the states $l \in \bar{l}$.
The normalization constant is denoted by $Z$, and $\bar{n}$ is a
vector containing the $n_{\bar{i}}$.
This formula is only approximate because it includes
configurations of the system that cannot occur in the simulations
since 
they are not connected by collisions with the initial 
configuration.
For small temperatures the sum in Eq.~(\ref{flu}) is readily
calculated numerically, and we will compare 
this with our simulation results in
Sec.~\ref{3D:particledistributions}.

\subsubsection{Collision time}

For any given configuration ${{\bf {n}}}$ of the system we calculate
the sum over 
all the transition matrix elements for collisions that can occur.
This sum
is the value of the right hand side of Eq.~(\ref{QBME}) for a given
${{\bf {n}}}$,
the corresponding $w_{{{\bf {n}}}}=1$ and all the other $w_{{{\bf
{m}}}}$ equal to zero.
A single possible collision $1+2 \rightarrow 3+4$ contributes to this
sum 
\begin{equation}
P(12\rightarrow34) = \frac{4 \pi}{\hbar \Delta e}
\left|
U_{1234}\right|^2n_{1}n_{2}\left(n_{3}+1\right)\left(n_{4}+1\right).
\label{traelem}
\end{equation}
where the factor of $4$ is due to different permutations of the
indices which
describe the same collision. We call a collision {\em possible} if it
conserves energy
and $U_{1234} \neq 0$, and we call $P(12\rightarrow34)$ the {\em
transition probability
per unit time} for this particular collision. 

\subsection{Ergodic approximation}
\label{ergass}

In solving the QBE or the classical Boltzmann equation it is a common
approximation to 
simplify this equation by an ergodic assumption
\cite{TraSimMurray,KinWal,TraSimQuantMurray,BECKagan,BECSemikoz}.
In a classical context this corresponds to the assumption that the
phase space density 
$f_t({{\bf {p}}},{{\bf {x}}})$ only depends on the energy $e$ of the
particles at position 
${{\bf {x}}}$ with momentum ${{\bf {p}}}$ at time $t$.
Quantum mechanically, it is postulated that degenerate energy levels
carry  equal populations at all times,
i.e.~the populations of degenerate eigenlevels equalize on a time
scale much faster
than collisions between levels of different energies.
This implies that 
the occupation numbers $n_i$ in the QBME should be replaced by
\begin{equation}
n_i \rightarrow n_{\bar{i}} = \frac{1}{g_{\bar{i}}} \sum_{i \in
\bar{i}} n_i,
\end{equation}
Here we define sets of indices $\bar{i}$ that contain all the indices
of one 
particles states with the same energy $\hbar \omega_i$; and
$g_{\bar{i}}$ is the 
degeneracy factor of states with 
energy $\omega_{\bar{i}}$. We note that the $n_{\bar{i}}$ are no
longer integers. 
In our simulation this corresponds to a distribution function which
is completely 
specified by the occupation numbers of (the block of) degenerate
energy levels, 
i.e. $w_{{{\bf {n}}}} \rightarrow w_{\bar{{{\bf {n}}}}}$
where $\bar{{{\bf {n}}}}$ is a vector containing the number of
particles in 
the degenerate eigenlevels $n_{\bar{i}}$. Removing or adding a
particle to
a state $\bar{i}$ changes $n_{\bar{i}}$ by $1/g_{\bar{i}}$.
Therefore, we use a vector ${{\bf {e}}}_{\bar{1}
\bar{2}\bar{3}\bar{4}}$ which is defined by 
\begin{eqnarray}
{{\bf {e}}}_{\bar{1} \bar{2} \bar{3} \bar{4}}& = &
[0,...0,\stackrel{\bar{1}}
{1/g_{\bar{1}}},0,...0,\stackrel{\bar{2}}{1/g_{\bar{2}}},0...
\nonumber \\
& & \quad ...0,\stackrel{\bar{3}}{-1/g_{\bar{3}}},0,...0,
\stackrel{\bar{4}}
{-1/g_{\bar{4}}},0,...0], 
\end{eqnarray}
to describe collisions in the ergodic case.

Using these definitions we can write the ergodic form of
Eq.~(\ref{QBME})
in the following way
\begin{eqnarray}
\dot{w}_{\bar{{{\bf {n}}}}} & = & -\frac{\pi}{\hbar}
\sum_{\bar{1}\bar{2}\bar{3}\bar{4}} 
\left( \sum_{\stackrel{1 \in \bar{1}, 2 \in
\bar{2}}{\scriptscriptstyle 3 \in \bar{3},
4 \in \bar{4}}} \delta\left(\hbar(\omega_{1} + \omega_{2} -
\omega_{3} - 
\omega_{4})\right) \left| U_{1 2 3 4} \right|^{2}\right)
\nonumber \\
&&\quad \times \left\{
n_{\bar{1}} n_{\bar{2}} \left( n_{\bar{3}}+1 \right) \left(
n_{\bar{4}}+1 \right)
\left[ w_{\bar{{{\bf {n}}}}}- w_{\bar{{{\bf {n}}}}+{{\bf
{e}}}_{\bar{1}\bar{2}\bar{3}\bar{4}}} \right] \right.
\nonumber \\
 & & \quad + \left. \left( n_{\bar{1}} +1\right) \left( n_{\bar{2}} +
1 \right)
n_{\bar{3}} n_{\bar{4}}
\left[ w_{\bar{{{\bf {n}}}}}- w_{\bar{{{\bf {n}}}}-{{\bf
{e}}}_{\bar{1}\bar{2}\bar{3}\bar{4}}} \right] \right\}
.\label{QBMEerg}
\end{eqnarray}
Transition probabilities $P(\bar{1}\bar{2}\rightarrow
\bar{3}\bar{4})$ are calculated according to  
\begin{equation}
P(\bar{1}\bar{2}\rightarrow \bar{3}\bar{4}) =
\sum_{\stackrel{1 \in \bar{1}, 2 \in \bar{2}}{\scriptscriptstyle 3
\in \bar{3}, 
4 \in \bar{4}}} P(12\rightarrow 34)
\label{ergtran}
\end{equation}
where the sum runs over all the elements of a particular set of
degenerate states. 
Collisions which do not change the energy distribution are thus no
longer taken into 
account. 

We note that the ergodic assumption yields the correct steady state
distribution, 
but we expect differences in the details of the dynamics.
A comparison of the kinetics with and without the ergodic assumption
will be given \
in the case of a 3D box in Sec.~\ref{3Dbox}; our simulation 
results for the harmonic oscillator in Sec.~\ref{HO} will be based on
the ergodic approximation.

\subsection{Comparison between the QBME and the QBE}
\label{CompQBMEQBE}

Recent work of kinetics in relation to Bose condensation in trapping
potentials by Holland and collaborators
\cite{TraSimMurray} is based on the QBE with an ergodic assumption
(for a classical Boltzmann equation see also 
\cite{KinWal}).  
The derivation of the QBE is based on factorizing mean values  
$
{\left\langle {n_1 n_2 ...n_i} \right\rangle}={\left\langle {n_1}
\right\rangle} {\left\langle {n_2} \right\rangle} ...{\left\langle {n_i} \right\rangle}
$
 with ${\left\langle {n_i} \right\rangle}=\sum_{{{\bf {n}}}} n_i
w_{{{\bf {n}}}}$.
In the ergodic approximation one obtains
\begin{eqnarray}
g_{\bar{1}} {\left\langle {\dot{n}_{\bar{1}}} \right\rangle}& = &
\frac{4 \pi}{\hbar} \sum_{\bar{2}\bar{3}
\bar{4}} \left\{-{\left\langle {n_{\bar{1}}}
\right\rangle}{\left\langle {n_{\bar{2}}} \right\rangle}({\left\langle {n_{\bar{3}}} \right\rangle}+1)
({\left\langle {n_{\bar{4}}} \right\rangle}+1) \right.+ \nonumber \\ 
& & \qquad \left. ({\left\langle {n_{\bar{1}}}
\right\rangle}+1)({\left\langle {n_{\bar{2}}} \right\rangle}+1){\left\langle {n_{\bar{3}}} \right\rangle}
{\left\langle {n_{\bar{4}}} \right\rangle}\right\} \nonumber \\  
& & \left(\sum_{\stackrel{1 \in \bar{1}, 2 \in
\bar{2}}{\scriptscriptstyle 3 \in \bar{3}, 
4 \in \bar{4}}} \left|U_{1234}\right|^2 \delta\left(\hbar(\omega_1 +
 \omega_2 - \omega_3 - \omega_4) \right)\right). 
\label{QBE}
\end{eqnarray}
In this equation ${\left\langle {n_{\bar{i}}} \right\rangle}$ is the
mean occupation of the degenerate states. 
The discrete ${\left\langle {n_{\bar{i}}} \right\rangle}$ replace
the particle distribution function $f(e)$ used in the classical
version \cite{KinWal}. 
The QBE describes the time evolution of the single particle
distribution function in the 
mean distribution of the other particles. In contrast to  simulations
of the QBME, 
fluctuations in the occupation numbers are thus not described  by
the QBE. Moreover, it is not possible to simulate systems far from 
equilibrium where the factorization of the mean values is not valid.
On the other hand, the QBE has the advantage that it allows
simulations with much larger particle numbers than the QBME.

\subsection{Simulation of the QBME}

Since the QBME is a stochastic master equation for (quantum
mechanical)
occupation probabilities, we can simulate its time evolution as a
series of
 jumps.
A jump describes the collision of two particles $1+2 \rightarrow
3+4$,
which is represented as an instantaneous change of the corresponding
occupation 
numbers. The simulation method is often used for rate equations and
works as follows:

\begin{enumerate}
\item Take an initial configuration of particles, $ {{\bf {n}}}$,
(representing an initial density operator
$\rho_{\ell}(t=0)={\left| {{{\bf {n}}}} \right\rangle}{\left\langle
{{{\bf {n}}}} \right|}$), where the energy
$E$ and the total number of particles $N$ are fixed.

\item Calculate all the transition probabilities per unit time
 $P(12\rightarrow34)$ for
the given ${{\bf {n}}}$.

\item The {\em total collision rate\/} is now proportional to the sum
over all 
transition probabilities per unit time.

\item The next jump occurs at time $t_m$ since the last jump, which
can be
calculated by choosing
a random number $r \in ]0,1]$ from a uniform distribution and using
\begin{equation}
t_{m} = -\frac{\ln(r)}{\sum_{1234} P(12\rightarrow34)}.
\label{meantime}
\end{equation}

\item All the possible collisions  are lined up with the length
$P(12\rightarrow34)$. Another random
number $s \in ]0,  \sum_{1234} P(12\rightarrow34)]$ is chosen from a
uniform 
distribution. The transition selected by this random number $s$ gives
the  particular collision 
$12 \to  34 $ which occurs.

\item The last step now is to set $t:=t+t_{m}$,
${\left| {{{\bf {n}}}} \right\rangle}:={\left| {{{\bf {n}}}-{{\bf
{e}}}_{1234}} \right\rangle}$ and
$\rho_{\ell}:={\left| {{{\bf {n}}}} \right\rangle}{\left\langle
{{{\bf {n}}}} \right|}$.

\item Go back to 2.

\item Repeat this simulation to obtain
$\rho= \mbox{constant} \times \sum_{\ell}\rho_{\ell}$.
\end{enumerate}

In every collision only four of the occupation numbers are changed
and
therefore only few of the transition matrix elements are modified by
the change 
in the occupation numbers.
Thus it is not necessary to calculate all transition probabilities
after each 
step, since only those involving the $ n_1$, $ n_2$, $ n_3$, $ n_4$
which
define the collision will have been changed.  (This is, however, more
complicated than is the case for the Boltzmann master equation, where
the
$ 1+n_i$ factors do not occur)

For integer occupation numbers it
is not possible to neglect the $ (1+n_i)$ factors above a certain
energy
by arguing that the mean occupation of highly excited states is much
smaller
than one. For highly excited states these factors are either $ 1$ or
$ 2$, {\em etc.} and cannot be replaced by $1$. We have to account
for them regardless of
the energy of the one-particle states involved into the collision.
In our simulation method we do not restrict the number states
available for
the particles of the gas. We keep track of each of the particles
rather than
of a certain number of one-particle states. This limits the number of
particles
we are able to consider.

\section{3D square well potential with periodic boundary conditions}
\label{3Dbox}

\subsection{Description of the system}

First, we will simulate the QBME for a 3D cube of length $L$ with  
periodic boundary conditions. This corresponds to the simplest
version of the QBME.
In the language of QKI $\delta x = L$ is the length of the phase
space cells. 
From this we immediately find the spacing of the cells in momentum to
be
$\delta p = 2 \pi \hbar / L$, which is equal to the
momentum spacing of the discrete energy levels in the box.
The wavelet functions (introduced in QKI Eq.~(26)) are therefore
reduced to
\begin{equation}
v_{{{\bf {k}}}}({{\bf {x}}})=\frac{\mbox{e}^{i {{\bf {k}}} {{\bf
{x}}}}}{L^{3/2}}
\end{equation}
We have dropped ${{\bf {r}}}$ of QKI in the equation above because
there is only one phase 
space cell
in coordinate space.
The wave numbers ${{\bf {k}}}$ take on the discrete values
\begin{equation}
{{\bf {k}}}=\frac{2 \pi}{L} {{\bf {m}}}
\end{equation}
where ${{\bf {m}}}$ is a vector consisting of integer values.
Since our system has the finite volume $L^3$ the wavelet functions
are
orthogonal in the following sense
\begin{equation}
\int\limits_{L^3} d^3\!x \: v_{{{\bf {k}}}_i}\! \left({{\bf
{x}}}\right) \:
v_{{{\bf {k}}}_l}^*\!\left({{\bf {x}}}\right) = {\delta_{{i},{l}}}
\end{equation}
With these wave functions we can now calculate $U_{1234}$ to be
\begin{equation}
U_{1234}=\frac{4 \pi \hbar^2 a}{m L^3} \:{\delta_{{{{\bf
{m}}}_1+{{\bf {m}}}_2},{{{\bf {m}}}_3+{{\bf {m}}}_4}}}.
\end{equation}

In the 
case of the 3D box $\Delta e = (2\hbar^2\pi^2)/(m L^2)$.
Using $\sigma=8\pi a^2$ \cite{TraSimQuantMurray} for the cross
section,
$\bar{n}=N/L^3$ and $v_1=(2\pi\hbar)/(mL)$ which is the
magnitude of velocity of a particle in the first excited state we get
\begin{eqnarray}
P(12\rightarrow34) &= &\sigma \bar{n} v_1 \frac{2}{N \pi}
\:{\delta_{{{{\bf {m}}}_1+{{\bf {m}}}_2},{{{\bf {m}}}_3+{{\bf
{m}}}_4}}} \nonumber \\
& & \quad n_1 n_2 \left(n_{3}+1\right)\left(n_{4}+1\right).
\end{eqnarray}
The number of possible collisions is restricted by two Kronecker
delta
functions that ensure energy and momentum conservation.
The overlap integral is $U_{1234} = (4 \pi \hbar^2 a)/(m L^3)$ for
all the possible collisions, and does not depend on energy
or momentum of the involved one-particle states.

In semi-classical treatments of the QBE, the ergodic assumption is
often made
\cite{BECKagan,BECSemikoz}.
The density of states is approximated to be proportional to
$\sqrt{e}$.
It is then shown that the transition matrix elements are proportional
to 
$\sqrt{e_{\rm min}}$, where $e_{\rm min}$ is the minimum energy of
the 
colliding particles (compare Appendix \ref{clcal}). In 
case of a smoothly varying, 
strictly decreasing function $f(e)$, one can therefore argue that
most
of the collisions  happen between particles with almost the same
energy.
In the cases we are interested in we cannot make these assumptions.
The occupation numbers can vary strongly and the degeneracy of states
which we count exactly is not proportional to $\sqrt{e}$ in the
energy
range in which our simulations are performed.

\subsection{Results of simulations}

All the simulations we report contain a statistical error. Unless
this
statistical error is given explicitly it is less than 5\%.

\subsubsection{Thermodynamic quantities}
\label{rebox}

There are two ways of computing the stationary solution of the QBME
(\ref{QBME}).
The first is to calculate it directly from Eq.~(\ref{statsol}); 
this is only feasible for very few atoms.
The second possibility is to obtain the stationary solution from
simulations
by assuming that the time average over a sufficiently long time
period  equals the 
ensemble average. To find this time we wait until the simulation
results agree with
a Bose-Einstein distribution Eq.~(\ref{BEDstat}).
This also allows us to assign a temperature $T$ to the system.
All the results are scaled to the critical temperature in the
thermodynamic limit $T_{c}=2\hbar^2\pi/(m L^2 k)
(N/\zeta(3/2))^{2/3}$
\cite{Pathria}. There are three parameters of the system;
$E, \: L$ and $N$ which give a certain $T, \: T_c$ and $N$ in 
thermodynamic equilibrium. In the simulation we fix $E$ and $N$, and
the scaling
to the critical temperature is equivalent to scaling to a certain
particle 
density $N/L^3$ in the box. 

The expression $T_c$ for the critical temperature is, of course, only
valid in the 
thermodynamic limit because in the derivation \cite{BECBagnato} sums
over energies are 
replaced by integrals which over- or underestimate the sums for
finite 
systems depending on the density of states. In the thermodynamic
limit the 
energy spectrum becomes continuous and summing yields the same result
as 
integrating. For a finite number of particles we therefore do not
expect
the critical temperature and the condensate fraction vs.~ temperature
to be the same as in thermodynamic limit.

Fig.~\ref{fiSQK1} shows the comparison of the results from the
simulations 
and the grand canonical expression Eq.~(\ref{grca}).
The degeneracies $g_{\bar{i}}$ in Eq.~(\ref{grca}) are calculated by
counting all
ways  of combining different integer numbers $m^x_i$, $m^y_i$,$m^z_i$
consistent with the definite energy
\begin{equation}
\hbar \omega_i = \frac{2 \pi^2 \hbar^2}{m L^2}
\left(\left(m^x_i\right)^2+
\left(m^y_i\right)^2+\left(m^z_i\right)^2\right).
\end{equation}

The simulation and the grand canonical result both give a higher
number of
particles in the condensate than expected from the thermodynamic
result.
The results for finite number of particles, however, approach 
the thermodynamic limit with increasing $N$ very quickly.
Around the critical temperature there is a slight deviation of the
simulated
results from the grand canonical results \cite{MicHolthaus}, whereas
for $T \ll T_c$ there 
is almost no difference.
This is due to a bigger statistical error in the simulation because
of the large fluctuations in the region around the critical
temperature.
Note also that we are comparing two different statistical ensembles, 
and that for finite systems we would not expect exact agreement
between
the results from different ensembles.

\subsubsection{Occupation of the ground state}
\label{ogs}

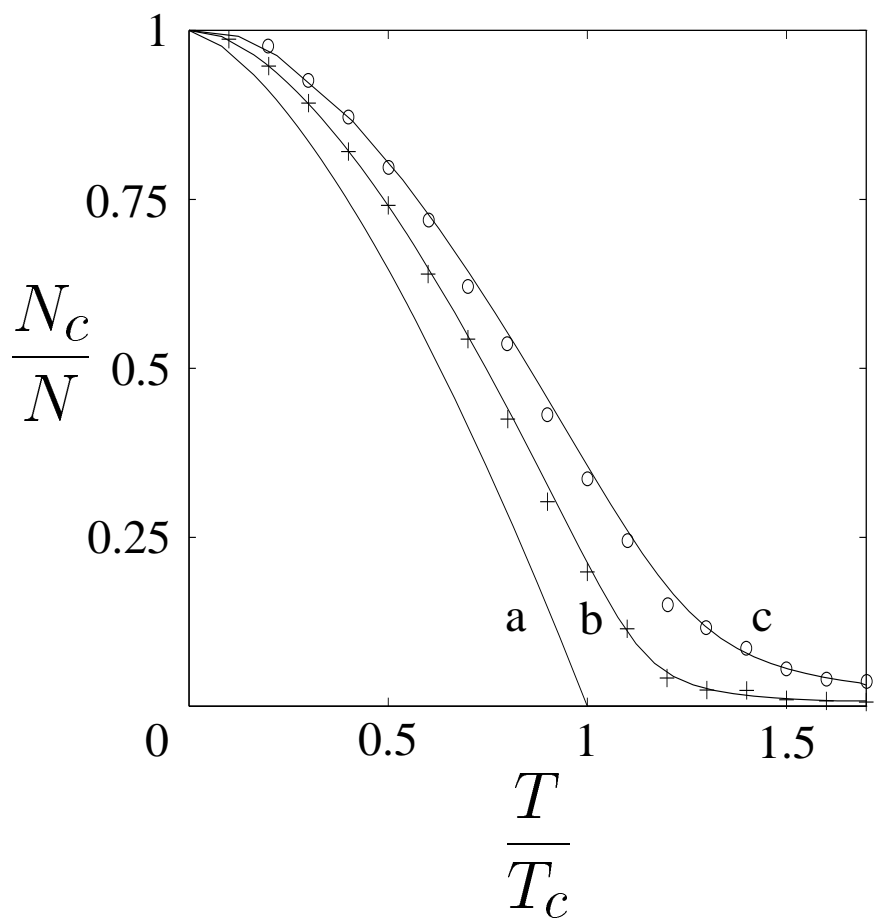
\begin{figure}
\begin{center}
\caption{Occupation numbers for the ground state $N_c$ and the first
excited 
state $n_1$ against the total number of particles $N$ in thermal
equilibrium 
for $T=0.5\:T_c$.\label{fiSQK10}}
\end{center}
\end{figure}

We want to investigate the scaling of the one-particle state
occupation with the
number of particles in the gas.
For BEC we expect \cite{Huang} the occupation of the ground state to
be
\begin{equation}
\frac{n_0}{L^3}=\frac{N}{L^3}\left(1-\left(\frac{T}{T_c}\right)^{3/2}\right),
\end{equation}
while for excited states 
\begin{equation}
\frac{{\left\langle {n_i} \right\rangle}}{L^3} \leq \frac{T}{L}
\times \frac{mk}{2 \pi^2 \hbar^2 {{\bf {m}}}_i^2} \quad
\stackrel{L \rightarrow \infty}{\longrightarrow} \quad 0.
\end{equation}
Fig.~\ref{fiSQK10} shows 
the occupation numbers of the ground state and the first excited
state.
At a given $T/T_c$ the number of particles in the ground state
increases
linearly with the total number, whereas the slope of the occupation
of the
first excited state becomes smaller with increasing number of
particles. 
From this numerical result we conclude that the QBME 
does really describe a macroscopic occupation and is consistent with 
expecting BEC below the critical temperature $T_c$.

\subsubsection{Collision times with and without the ergodic
assumption}

Obviously, the results given in previous Secs.~\ref{rebox} and
\ref{ogs} 
are the same with or without use of the ergodic
assumption. This is expected, because in thermal equilibrium all
degenerate
eigenlevels should have the same occupation number even without  the
ergodic
assumption. The mean time between two collisions in
thermal equilibrium is computed by taking the  average over all the
calculated 
times $t_m$ from Eq.~(\ref{meantime}). This time has to be multiplied
by $N/2$ because 
one particular particle is involved in one out of $N/2$ collisions. 

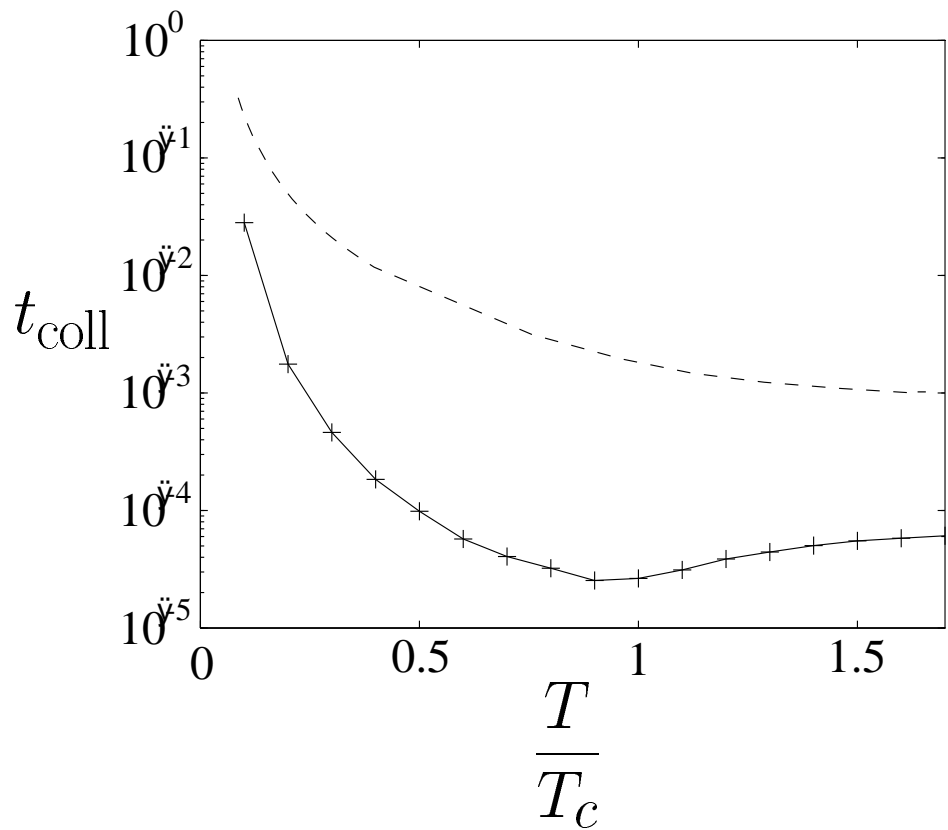
\begin{figure}
\begin{center}
\caption{Mean collision time per particle versus temperature for
$N=100$ for 
the
3D box. Result of
the simulation without (b) and with (a) the ergodic assumption,
result for $t_{\rm coll}=(\sigma \bar{n} v_{T})^{-1}$ (dashed line).
The
time scale is normalized to $(\bar{n} \sigma v_1 2/N)^{-1}$. $v_1$ is
the magnitude of
velocity of a particle in a first excited state as defined in the
text.\label{fiSQK9}}
\end{center}
\end{figure}

In Fig.~\ref{fiSQK5} we plot the mean collision time $t_{\rm
coll}^{\rm ne}$
for one particle versus the temperature without the ergodic
assumption .
The classical elastic mean collision time calculated from $t_{\rm
coll}^c=
(\sigma \bar{n} v_T)^{-1}$, where $v_T$ is the mean thermal velocity
of the gas $v_T=N^{-1} \sum_i \sqrt{(2 \hbar \omega_i)/(m)}
{\left\langle {n_i} \right\rangle}$,
and ${\left\langle {n_i} \right\rangle}$ is the mean occupation of
the i-th energy level obtained from 
the simulation. As soon as the gas becomes degenerate the $1+n_i$
factors in 
Eq.~(\ref{QBME}) become important and increase the collision rate
compared to the 
classical case. For temperatures close to zero the collision time
increases again because there are only few particles outside the
condensate
which can take part in collisions. Fig.~\ref{fiSQK9} shows the
comparison 
between the curves for the ergodic collision time $t_{\rm coll}^{\rm
e}$ 
and $t_{\rm coll}^{\rm ne}$. For very small temperatures, the ergodic
assumption 
allows for collisions which can not occur in the non-ergodic case
because the 
corresponding states are not occupied. As soon as the temperature is 
close to $T_c$, those collisions in the non-ergodic case that only
change the 
direction of the momentum of an individual particle decrease $t_{\rm
coll}^{\rm ne}$ 
compared to the ergodic case. Since this type of collisions leaves
the energy of the
particles unchanged they are not included in the ergodic
calculations. 

\subsubsection{Particle distributions}
\label{3D:particledistributions}

While the mean values for the occupation numbers are easy to
calculate it takes
more effort to find the particle distribution $W_i(j)$ of the
one-particle states.
There are two ways of calculating these distributions.
Either we calculate the time a state was occupied by a certain number
of
particles (time average) or we record the number of particles in that
state after a certain time for many different trajectories (ensemble
average).
Both results need not necessarily be the same 
unless the system has the stationary solution Eq.~(\ref{statsol}).
We used both methods to calculate particle distributions for the
condensate and some of the excited states for different temperatures.

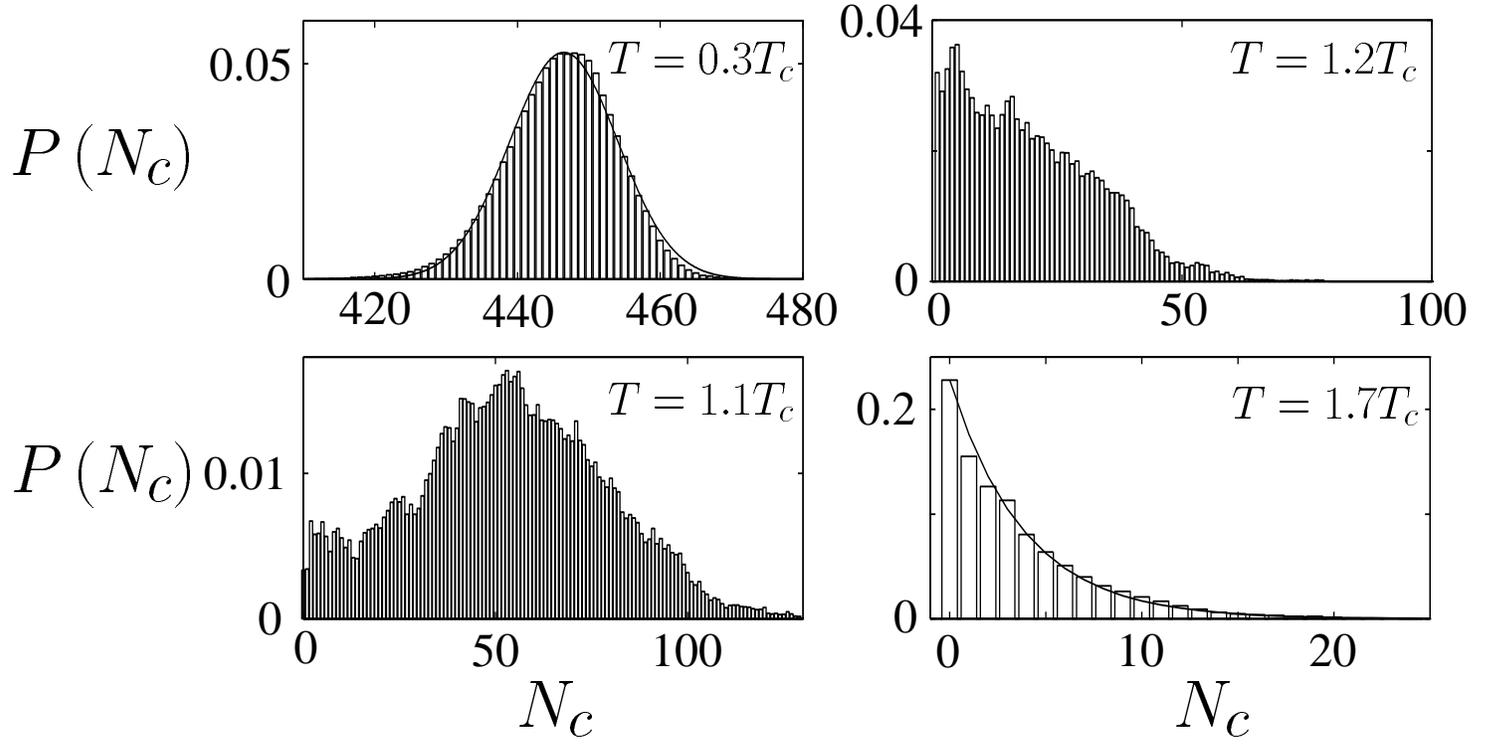
\begin{figure}
\begin{center}
\caption{Probability distribution of particles in the condensate for
the
3D harmonic box without the ergodic assumption. Results of the
simulation (bars) and fits (solid lines) are calculated
for $N=500$. \label{fiSQK3}}
\end{center}
\end{figure}

In Fig.~\ref{fiSQK3} particle distributions for the ground state are
plotted.
Particle distributions of the condensate are well approximated by a
Gaussian for
temperatures below $T_c$.
However, they are not completely symmetric around the mean value like
the  
Gaussian there is a slight asymmetry which increases with
temperature. 
The shape of the distribution changes close to the critical
temperature.
For $N=500$ at $T=1.1T_c$ the distribution develops a second local
maximum at  
$N_c=0$ and
at $T=1.2T_c$ the peak at finite number of condensate particles has
disappeared. 
Well above $T_c$ at $T=1.7 T_c$ it agrees with a Bose-Einstein
distribution
\begin{equation}
p(N_c)=(1-\eta)\eta^{N_c}
\label{BEDdi}
\end{equation}
with ${\left\langle {N_c} \right\rangle}=\eta/(1-\eta)$.

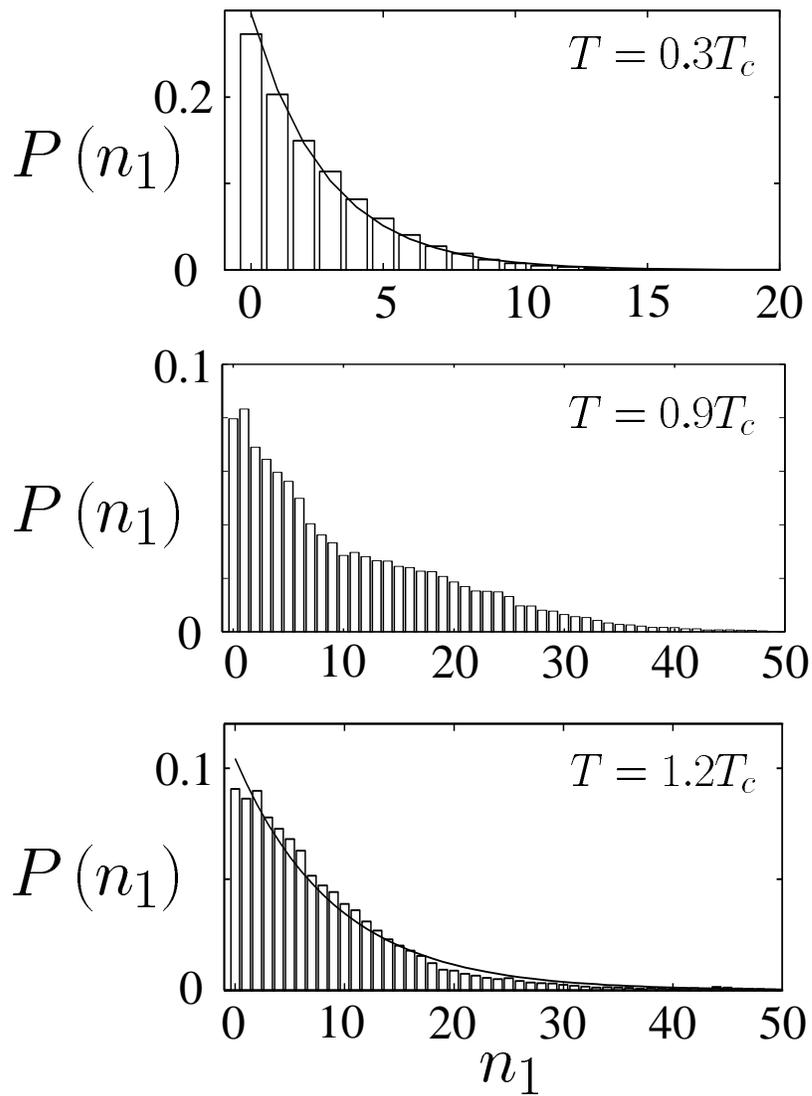
\begin{figure}
\begin{center}
\caption{Probability distribution of particles in one of the
first excited states for the
3D box without the ergodic assumption. Results of the
simulation (bars) and fits (solid lines) are calculated
for $N=500$. \label{fiSQKJ}}
\end{center}
\end{figure}

The particle distribution of the first excited state in
Fig.~\ref{fiSQKJ}
can be approximated by the Bose-Einstein distribution (\ref{BEDdi})
for 
$T \ll T_c$ and $T \gg T_c$. 
Particle distributions of highly excited states agree with the  
Bose-Einstein probability distribution (\ref{BEDdi}) at all
temperatures.

In Fig.~\ref{fiSQK2} we plot the standard deviation $\sigma(N_c)$ of
the 
particle distribution of the condensate in thermal equilibrium.
The error bars are calculated according to
$\sqrt{\sigma((N_c-{\left\langle {N_c} \right\rangle})^2)/{\left\langle {N_c} \right\rangle}}$
which is the variance of the standard deviation normalized to the
mean number
of particles in the condensate. This gives the mean deviation of
the standard deviation from its calculated value.
For small temperatures $\sigma(N_c)$ rises almost linearly with
temperature.
The number of possible states with different number of particles in
the
condensate increases which leads to a larger width of the
distribution. 
Close to $T_c$ we get a very broad particle distribution with a very
large
standard deviation. 
For $T>T_c$ the standard deviation will tend to go to the mean number
of particles
in the condensate which agrees with the fit to the Bose-Einstein
distribution
which has a standard deviation of $ N_c(N_c+1)$ going to $ N_c$ for
$N_c \ll 1$.

This also agrees with the calculation performed in Sec.~\ref{partdis}
for the
case of very small mean occupation of a state.
At small temperatures we calculate the particle distribution in the
condensate
according to Eq.~(\ref{flu}). To compute the sum in Eq.~(\ref{flu})
we assume that most of the fluctuations come from exchange of
particles
of the condensate with the first few excited states. The particles
in higher excited states should not have a significant influence on
the fluctuations
in the condensate, but they should ensure that particles in the
lowest lying states
can be distributed among degenerate eigenlevels, without restrictions
due
to conservation laws. The derivation of Eq.~(\ref{flu}) is based on
the 
assumption that the particle distribution among degenerate
eigenlevels is not
restricted by conservation laws. Using Eq.~(\ref{flu}) 
we obtain particle number fluctuations of the condensate due to
exchange with low lying levels. 
In particular we calculate $W_0(j)$ from Eq.~(\ref{flu}) by taking
into account the
first seventeen energy levels. The particle distributions we get
agree well
with the ones from the simulations. In Fig.~\ref{fiSQK2} the results
of both
calculation methods are compared for temperatures $T < 0.5 T_c$.
The crosses correspond to the numerical calculations on
Eq.~(\ref{flu}) and 
agree well with the simulation results.

\subsubsection{Growth of the condensate}

Here we want to investigate how the condensate builds up when the
simulation
is started in a non-equilibrium distribution. As the initial state we
choose a Gaussian-like distribution: We first distribute the
particles
randomly into states with energies between that of the first excited
state,
and twice the mean energy, and then
move particles to higher or lower energy states until the given 
fixed energy $E$ of the system is exactly reached. 

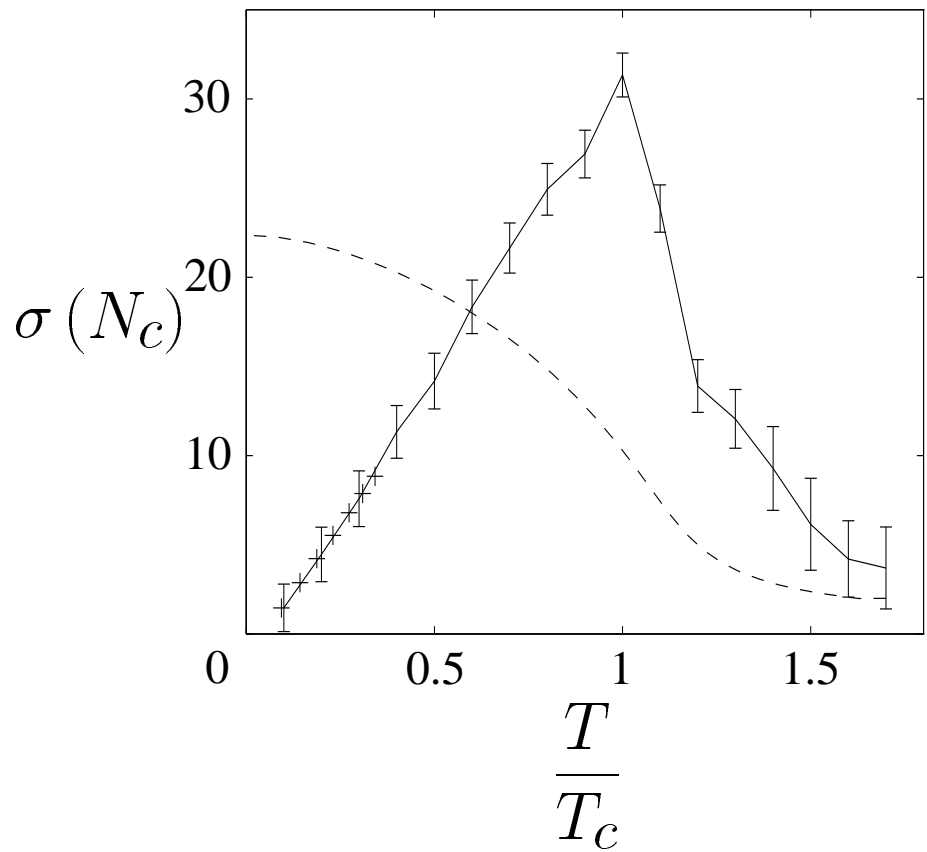
\begin{figure}
\begin{center}
\caption{Buildup of the condensate for the 3D box for $N=500$.
The energy is chosen such that in equilibrium $T=0.5T_c$. The
time scale is normalized to $(\bar{n} \sigma v_1 2/N)^{-1}$ in
thermal equilibrium. 
The dashed line is a fit of the form $N_c(1-\mbox{exp}(-t/\tau))$,
with
$\tau=0.0013$ and $N_c=368$, as explained in the text.
\label{fiSQK4}}
\end{center}
\end{figure}

Whenever possible we avoid putting particles in the condensate at the
beginning 
of the simulation. As can be seen in Fig.~\ref{fiSQK4} the condensate
growth is well fitted by $N_c\left(1-\exp(-t/\tau)\right)$ 
where $N_c$ is the number of particles in the condensate in thermal
equilibrium and the time constant $\tau$ is found by fitting this
function
to the simulation. This holds as long as the fraction of condensate
particles
in thermal equilibrium is not much less than one.

\subsubsection{Time to reach an ergodic distribution}

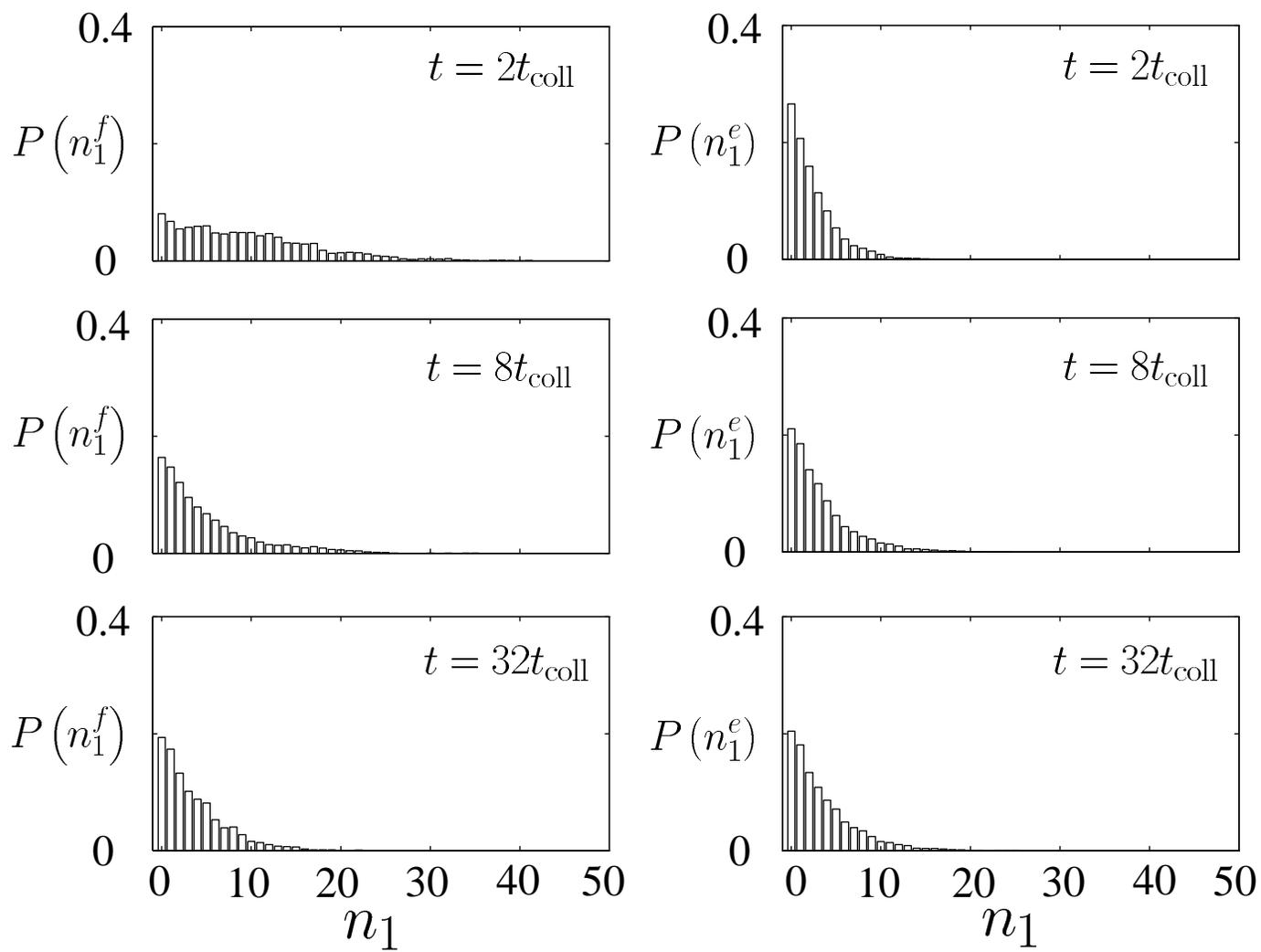
\begin{figure}
\begin{center}
\caption{Distortion of an ergodic distribution into a
non ergodic one for the 3D box. Particle distribution of
the depleted levels ($P(n_1^e)$) and of the filled
levels ($P(n_1^f)$) at time $t$ after the distortion.
Simulation for $N=500$ at $T=0.4T_c$.\label{fiSQKA}}
\end{center}
\end{figure}

While with the ergodic assumption all degenerate levels are equally
occupied at all times, in the non-ergodic case collisions themselves
are 
responsible for  equalizing the occupations of degenerate levels.
To check the relaxation time for a distribution to become ergodic we 
disturb a system in thermal equilibrium by putting {\em all}
particles 
with energy $\Delta e$ into {\em two} of the first excited states
(with 
opposite momentum so that the total momentum is unchanged).
As is shown in Fig.~\ref{fiSQKA} the particle distribution comes 
to equilibrium in approximately  $10 \: t_{\rm coll}^{\rm ne}$.
Collisions, therefore,
transfer the occupation between degenerate levels at a time scale of
the order 
of the mean collision time in the gas.
We conclude that for the ergodic assumption to be valid strictly
speaking it is 
only reasonable to look at quantities that 
are mean values over several collision times.

\section{3D isotropic harmonic oscillator}
\label{HO}

\subsection{Description of the system}

In this section we will study Bose particles trapped in an isotropic
harmonic 
trap with trap frequency $ \omega$. The vector ${\left| {{{\bf {n}}}}
\right\rangle}$ now gives 
the occupations of the trap levels, 
and $U_{1234}$ contains the spatial eigenfunctions of the harmonic
oscillator. 
For the low lying levels these integrals can be evaluated numerically
but
for highly excited states it is difficult to get reliable results for
$U_{1234}$. Therefore, we will limit ourselves to using the ergodic
form 
of the QBME as explained in Sec.~\ref{ergass}. As is shown in 
\cite{TraSimQuantMurray}, the transition matrix elements of
transitions which 
change the energy distribution function can be approximated by
\begin{eqnarray}
P(\bar{1} \bar{2} \rightarrow \bar{3} \bar{4}) & = & \frac{4
\pi}{\hbar^2
\omega}
\frac{m \sigma \omega^3 \hbar}{4 \pi^3}
g_{\mbox{min}(\bar{1} \bar{2} \bar{3} \bar{4})}^h \nonumber \\
& & n_{\bar{1}} n_{\bar{2}} (n_{\bar{3}}+1)(n_{\bar{4}}+1)\nonumber
\\
&=& \bar{n}_h^0 \sigma v_0 (2 N)^{-1} g_{\mbox{min}(\bar{1} \bar{2}
\bar{3} \bar{4})}^h  \nonumber \\
& & n_{\bar{1}} n_{\bar{2}} (n_{\bar{3}}+1)(n_{\bar{4}}+1)
\end{eqnarray}
Here $g_{\bar{j}}^h=(j+1)(j+2)/2$ is the degeneracy factor of the
$j$-th eigenstate with
energy $j\hbar \omega$,
$v_0=\sqrt{(4 \hbar \omega)/(\pi m)}$ the mean magnitude of velocity 
of a particle in the ground state of the oscillator, and
$\bar{n}_h^0$ 
is the mean particle density if all the particles are within a cube
of
length $\sqrt{(\hbar \pi)/(m \omega)}$.
This is the semi-classical expression obtained in Appendix
\ref{clcal}. 
According to Ref. \cite{TraSimQuantMurray}  numerical calculation 
shows
that the this expression is a good approximation even for low lying
energy levels.

\subsection{Results of simulations}

\subsubsection{Stationary solutions}

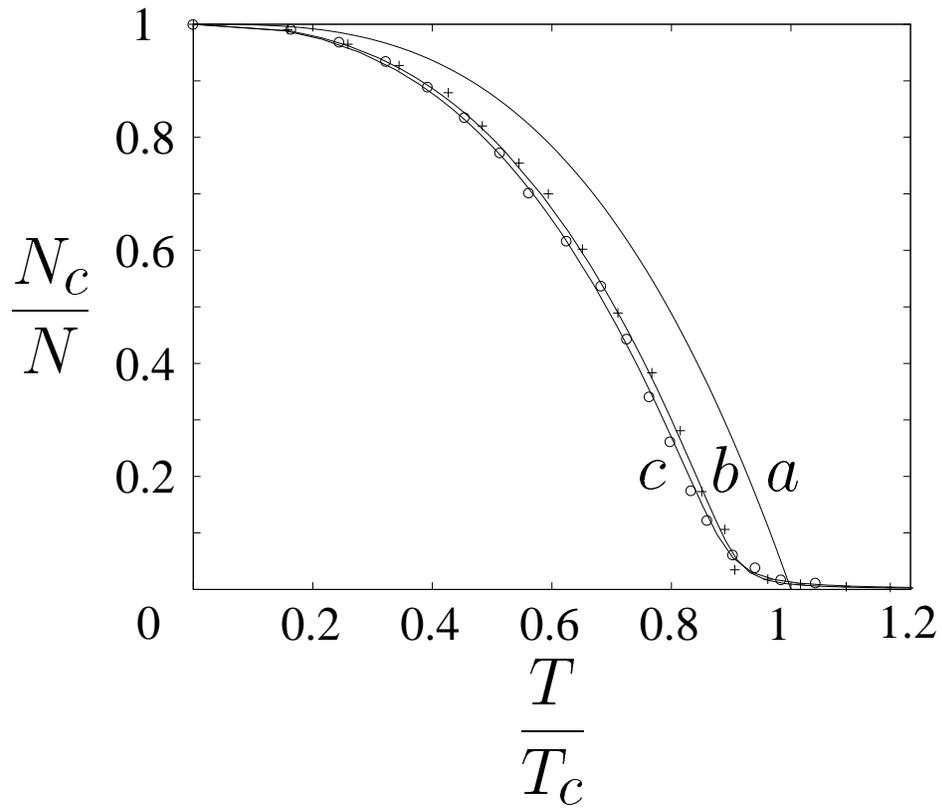
\begin{figure}
\begin{center}
\caption{Total energy of the system versus temperature. (a) data for
harmonic
oscillator, (b) data for the 3D box each with $N=500$.
Energy is normalized to the level spacing $\Delta e$. (+) results
from the
microcanonical simulation; (solid line) result of the grand canonical
calculation.\label{fiSQK7}}
\end{center}
\end{figure}

To obtain the grand canonical stationary solutions for the 3D
harmonic 
oscillator we have to replace $g_{\bar{i}}$ by 
$g_{\bar{i}}^h$ in the Eq.~(\ref{grca}). The critical temperature for
an ideal Bose gas in a 3D 
isotropic trap in the thermodynamic limit (i.e.~when the sums over
the
discrete energy levels are replaced by integrals) is given by 
$T_{c}=(\hbar \omega)/k \left(N/\zeta(3)\right)^{1/3}$ 
\cite{BECBagnato}. Our simulation results for the condensate fraction
versus temperature 
are shown in Fig.~\ref{fiSQK6}. The continuum approximation increases
the 
condensate fraction for finite number of particles compared to the
simulation
results. The reason for this is
that the density of states rises much faster than for the 3D box. 
As in the case of the square well potential, the results for the
microcanonical 
simulations and the grand canonical calculations agree very well.
Comparing the
two curves for $N=500$ of Fig.~\ref{fiSQK1} and Fig.~\ref{fiSQK6} we
find that the 
phase transition is more pronounced 
in the harmonic oscillator compared to the much smoother transition
for the 3D
box. This behavior can also be seen by plotting of the energy versus
temperature in Fig.~\ref{fiSQK7}. There is a visible change in the
slope of 
the energy for the harmonic oscillator even for $N=500$. For the
ideal gas the 
heat capacity  has a jump at the critical temperature in the
thermodynamic limit 
in the harmonic oscillator whereas in case of the 3D box only the
slope of the
heat capacity is discontinuous at the critical temperature
\cite{BECBagnato}
This makes clear that 
the thermodynamically expected differences in the condensation
process
between the harmonic oscillator and the free gas can also be seen in
finite
systems for small particle numbers. 

\subsubsection{Collision times}

We will now compare the mean collision time obtained in our
simulations 
$t_{\rm coll}^{\rm he}$ with the elastic collision time defined as
$t_{\rm coll}^{\rm hc}=(\bar{n}_h \sigma v_{\rm Th})^{-1}$.
We determine $v_{\rm Th}$ and $\bar{n}_h$ with the assumption 
that the  kinetic energy of the particles are equal
potential energy equally. Then we find for the mean density and the
thermal velocity 
$\bar{n}_h = (3N)/(4\pi)\left((m \omega^2)/(E_{3/2})\right)^{3/2}$
and
$v_{\rm Th}=\sqrt{E_{1/2}/m}$ respectively, with $E_s=\left(
1/N \sum_{i} \left( \hbar \omega_i \right)^s {\left\langle {n_i}
\right\rangle}\right)^{1/s}$.

In Fig.~\ref{fiSQK8}  we plot the mean collision time versus
temperature. 
For temperatures higher than the critical temperature the simulation
agrees well with
the classical result (dashed curve). For temperatures far below
$T_c$,
the result of the simulation is approximately equal
to the dotted curve which we obtained by the assumption that the size
of the cloud is
the ground state size and only the thermal velocity varies with
temperature.
Around the critical temperature the size of the cloud shrinks faster
than 
expected from the classical approximation.

\subsubsection{Evaporative cooling}

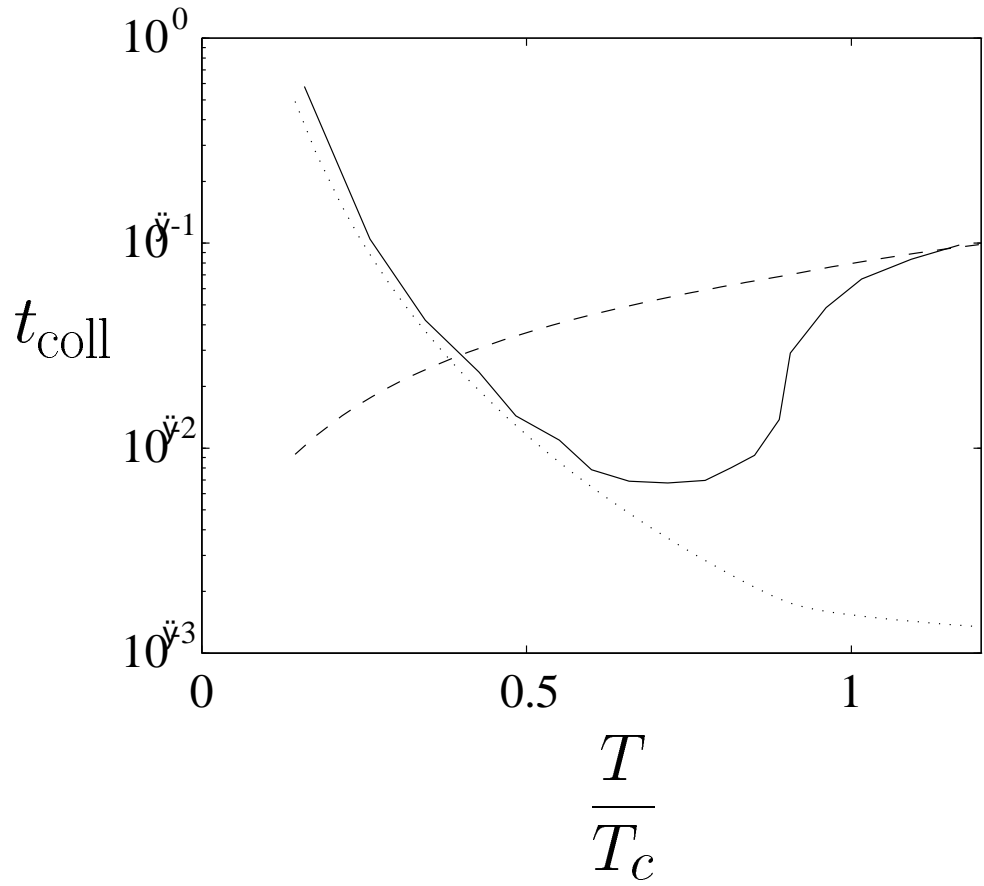
\begin{figure}
\begin{center}
\caption{Total number of particles $N$ and number of particles in the
condensate $N_c$ for $\gamma=1/10$ (solid line) $\gamma=1/2$ 
(dashed line) and $\gamma=3/2$ (dotted line) against time $t$.
$\gamma$ is the time constant from Eq.~(\protect\ref{board})
normalized to 
$\sigma \bar{n}_h^0 v_0 2/ N$. The time $t$ is normalized to 
$(\sigma \bar{n}_h^0 v_0 2/ N)^{-1}$. \label{fiSQKD}}
\end{center}
\end{figure}

Currently BEC is achieved in experiments by evaporative cooling,
i.e.~by removing particles
with a high energy from the trap (for a review see
\cite{evapKetterle}). Elastic collisions between the particles 
thermalize the particle distribution which leads to a decreasing
temperature.
To simulate a Bose gas that is evaporatively cooled we cut off the
trap
at a certain energy level $E_b(t)$, with $E_b(t)$ a given function of
time. Each particle which is scattered into an energy 
level above $E_b(t)$ after a collision is considered as lost. In our
simulations we
start with $N_0=800$ particles in the thermodynamic equilibrium at a
temperature of
$T \approx 1.4 T_c$. Then all particles with an energy larger than
$E_b(t=0)=65 \hbar
\omega$ are removed. During the simulation we decrease $E_b$
exponentially
according to 
\begin{equation}
E_b(t)=(E_b(0)-E_l) \mbox{e}^{-\gamma t}+E_l,
\label{board}
\end{equation}
where $E_l=8 \hbar \omega$. In Fig.~\ref{fiSQKD} the total number of
particles in
the gas $N$ and the number of particles in the condensate $N_c$ are
plotted as a function of 
time for different parameters $\gamma$. First the particles in the
highest energy levels
are evaporated quickly. During the cooling process, the collision
time decreases by an order
of magnitude as shown in Fig.~\ref{fiSQKE}; nevertheless the number
of particles evaporated per unit time
does not increase during the cooling process. The reason is that most
of the collisions
occur between particles with almost the same energy and thus many
collisions are necessary 
to redistribute the particles when some of them are evaporated. If
the collision rate did not 
increase so rapidly particles might be lost from the trap faster than
evaporative cooling is possible.
As soon as the condensate builds up the mean collision time increases
again. This 
expected behavior agrees qualitatively with Fig.~\ref{fiSQK8}. 

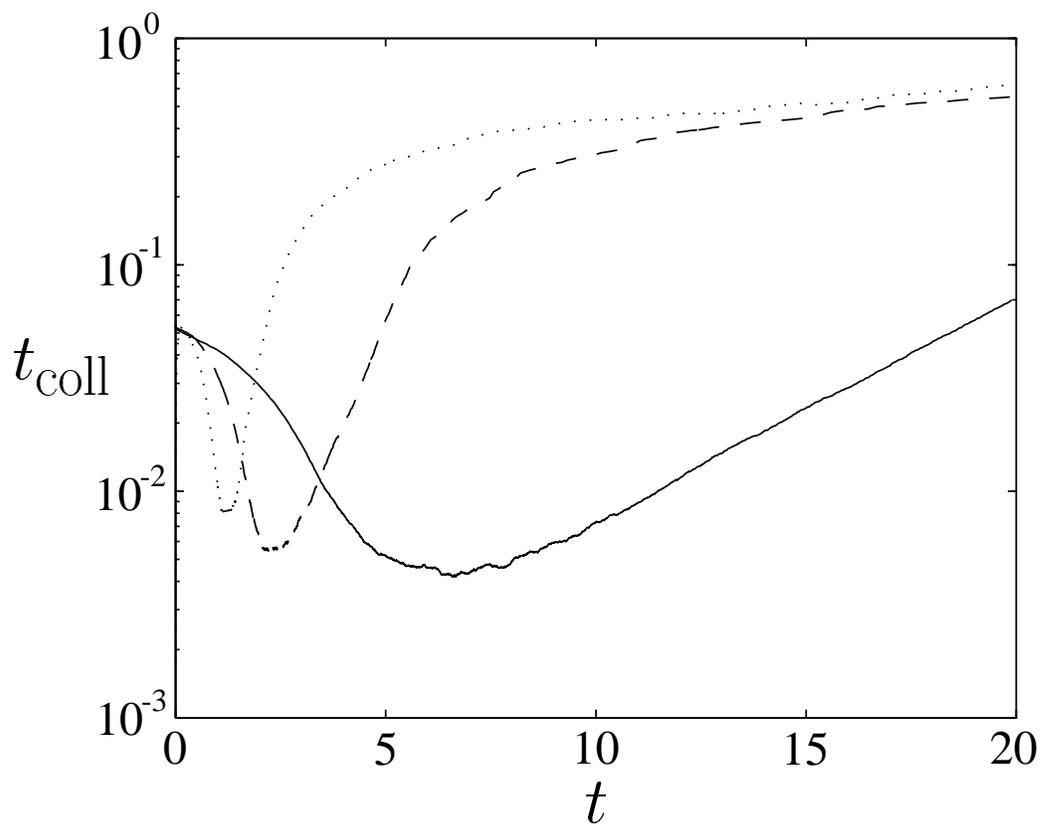
\begin{figure}
\begin{center}
\caption{Size of condensate divided by time to reach 90\% of the
final size
of the condensate versus time constant $\gamma$. $\gamma$ is the time
constant 
from Eq.~(\protect\ref{board}) normalized to 
$\sigma \bar{n}_h^0 v_0 2 / N$.\label{fiSQKF}}
\end{center}
\end{figure}

In order for the evaporative cooling to be efficient it is important
to quickly put as many particles as possible into the condensate. We
therefore have calculated
the size of the condensate divided by the time needed to reach $90\%$
of the equilibrium condensate fraction for 
different values of $\gamma$ as shown in Fig.~\ref{fiSQKF}. The size
of the condensate is limited by the initial number of particles in
the gas, the initial
size of the cut-off $E_b(0)$, and the initial energy $E$ (for $\gamma
\ll t_{\rm coll}^{-1}$).
For $\gamma \geq t_{\rm coll}^{-1}$ only few collisions will occur
while the cut-off is ramped down.
The number of particles that reach the condensate is therefore mainly
determined by 
the collision rate. As can be seen from the Fig.~\ref{fiSQKF} there
is a
value for the ramp rate $\gamma$ which maximizes the number of
particles transferred into the condensate per
unit time, and therefore optimizes the cooling process under the
assumption that additional loss
rates from the trap do not change while the gas is cooled.

We also performed some evaporative cooling simulations for a gas in a
3D-box. 
Because the density
of states in that case does not rise as quickly as in the harmonic
oscillator particle
energies are changed more during a collision than in the harmonic
oscillator. Therefore
it is also possible to evaporatively cool a gas in a box quickly
although the collision
rate does not rise as much as in the harmonic trap. 
 
\section{Conclusions}

We have simulated stationary and non-stationary properties of a Bose
gas in a trapping potential with finite number of particles in the framework of the Quantum Boltzmann Master Equation.

For a gas confined in a 3D-box we have found that
the number of particles in the condensate at a given temperature
is larger than expected from the thermodynamic limit.
We have also computed the mean collision time of particles
in the gas. Comparison with the classical result shows that
boson statistics tends to decreases the collision time close to the
critical temperature.
Calculations of fluctuations in the number of particles in the
one-particle
ground state have shown that the standard deviation increases almost
linearly with
temperature until the critical temperature is reached. For
temperatures above
$T_c$ the standard deviation decreases again,
and the distribution becomes Poissonian for high temperatures.
We have also found that population ´ is transferred 
at a time scale of the order of the collision time which is important
for the 
range of validity of the ergodic form of the QBME.

Our simulations of a Bose gas in an isotropic harmonic trap were
restricted to  the ergodic form of the QBME. 
In contrast to the 3D box the number of particles in the
condensate is decreased relative to the usual continuum
(thermodynamic) limit at a given
temperature. We found that the mean collision time decreases 
significantly as temperature reaches the critical point from above.
This is due to the increase of the density, as soon as the ground
state is macroscopically occupied.
Simulations of evaporative cooling have shown that there is an 
ramp rate to lower the cut-off energy of the trap with the goal of
transferring 
as many particles as possible per unit time to the ground state.

The present formalism is readily extended  to include mean field
effects, and pumping 
and loss of particles from a degenerate Bose gas. This is relevant
for modeling atom 
lasers based on collisions \cite{ALHolland,Wiseman}.

Acknowledgment:
D.J. thanks JILA for hospitality, supported by a fellowship from
the University of Innsbruck, and for very stimulating and useful
discussions with M.~Holland, and J.~Cooper.
D.J. and P.Z. were supported in part by the Austrian Science
Foundation.
C.W.G. is supported by The Marsden Fund.
Part of this work was supported by TMR network ERB 4061 PL 95-0044.

\appendix

\section{The classical limit}
\label{clcal}

To connect the present paper with Ref. \cite{KinWal} we briefly
rederive 
the classical Boltzmann equation with the ergodic approximation from
the QBE (\ref{QBE}). 
We assume the distance between energy levels small compared to the
mean
energy of a particle so that the sum can be replaced by an integral.
In the classical limit we get for the density of states at energy $e$

\begin{equation}
\rho(e)=\frac{1}{(2 \pi\hbar)^3} \int \, d^3p \, d^3x \,
\delta\left(e-U({{\bf {x}}})-
\frac{{{\bf {p}}}^2}{2m}\right),
\end{equation}
where $U({{\bf {x}}})$ is the trapping potential.
The degeneracy of the coarse grained one-particle states $g(e)$ is
connected to
the density of states by  $g(e)=\Delta e \rho(e)$.
We replace
\begin{eqnarray}
\sum_{i \in \bar{i}} \Psi^{\ast}_i({{\bf {x}}}) \Psi_i({{\bf {x}}}')
&\rightarrow&  
\frac{\Delta e}
{(2\pi\hbar)^3} \int d^3p_i \, \delta\left(e_i-U({{\bf
{x}}})-\frac{{{\bf {p_i}}}^2}{2m}
\right)  \nonumber \\
& & \qquad e^{i({{\bf {x}}}-{{\bf {x}}}'){{\bf {p}}}_i/\hbar},
\label{sumrepl}
\end{eqnarray}
where $e_i=\hbar \omega_i$. The factor $\Delta e$ in
Eq.~(\ref{sumrepl})
ensures the normalization of the sum over the wavefunctions to
$g(e_i)$.
Inserting replacement Eq.~(\ref{sumrepl}) into
$\left|U_{1234}\right|^2$ and integrating over ${{\bf {x}}}'$
yields a $\delta$--function of the four momenta times $(2 \pi
\hbar)^3$. Integrating 
over ${{\bf {p}}}_4$ i.e. setting ${{\bf {p}}}_4={{\bf {p}}}_1 +
{{\bf {p}}}_2 - {{\bf {p}}}_3$ we obtain
\begin{eqnarray}
\lefteqn{\sum_{\stackrel{1 \in \bar{1}, 2 \in
\bar{2}}{\scriptscriptstyle 3 \in \bar{3}, 4 \in \bar{4}}}
 \left|U_{1234}\right|^2 \delta(e_1+e_2-e_3-e_4) =} \hspace{0.5cm}
\nonumber \\
& &  \frac{16 \pi^2 \hbar^4 a^2}{m^2}
\frac{\Delta e^4}{(2\pi\hbar)^{9}} \int d^3p_1 \, d^3p_2 \, d^3p_3
\nonumber \\
& & \prod_{i=1}^{4} \delta\left(e_i-U({{\bf {x}}})-\frac{{{\bf
{p}}}_{i}^2}{2m}\right)
\delta(e_1+e_2-e_3-e_4).
\end{eqnarray}
We define the total momentum ${{\bf {P}}}={{\bf {p}}}_1+{{\bf
{p}}}_2$ and the
relative momenta ${{\bf {q'}}}=({{\bf {p}}}_1-{{\bf {p}}}_2)/2$ and
${{\bf {q}}}=
({{\bf {p}}}_3-{{\bf {p}}}_4)/2$.
Integrating over the azimuthal angels of the two relative momenta
${{\bf {q}}}$ and ${{\bf {q'}}}$ and over 
the length of the relative momentum ${{\bf {q}}}$ and calculating the
remaining integral 
similarly to \cite{KinWal} we obtain
\begin{eqnarray}
\lefteqn{\sum_{\stackrel{1 \in \bar{1}, 2 \in
\bar{2}}{\scriptscriptstyle 3 \in \bar{3}, 
4 \in \bar{4}}} \left|U_{1234}\right|^2 \delta(e_1+e_2-e_3-e_4) =} 
\hspace{0.5cm}\nonumber \\
& & \frac{2 m \Delta e^4 a^2}{\pi^2 \hbar^2} \rho(e_{\rm min})
\delta(e_1+e_2-e_3-e_4).
\label{semirepl}
\end{eqnarray}
We insert expression (\ref{semirepl}) into the QBE (\ref{QBE}),
divide
by $\Delta e$, replace the notation ${\left\langle {n_{\bar{i}}}
\right\rangle}$
by $f(e_i)$ and the $\sum \Delta e^3$ by $\int de_2 \, de_3 \, de_4$
and obtain finally
\begin{eqnarray}
\rho(e_1) f(e_1)&=&\frac{8 \pi m a^2}{\pi^2 \hbar^3} \int de_2 \,
de_3 \, de_4
\delta(e_1+e_2-e_3-e_4) \nonumber \\
& & \left\{- f(e_1) f(e_2) +f(e_3) f(e_4) \right\} 
\rho(e_{\rm min}).
\label{BE}
\end{eqnarray}
In Eq.~(\ref{BE}) the $(1+f)$ factors are neglected by assuming that
in the classical limit
the mean occupation of a quantum level is much less than one. Setting
$\sigma=8 \pi a^2$ 
this is the ergodic form of the classical Boltzmann equation from
Ref. \cite{KinWal}.

\begin{figure}
\begin{center}
\caption{Condensate fraction versus temperature in thermal
equilibrium for
the 3D square well potential.
(a) thermodynamic limit, (b) grand canonical solution for $N=500$
(solid line)
and results from the simulation (+), (c) grand canonical solution for
$N=100$ (solid line) and results from the simulation
(o).\label{fiSQK1}}
\end{center}
\end{figure}

\begin{figure}
\begin{center}
\caption{Mean collision time per particle versus temperature for
$N=500$ of the
3D box without ergodic assumption. Result of
the simulation (+) and result for $t_{\rm coll}=(\sigma \bar{n}
v_{T})^{-1}$ 
(dashed line). The
time scale is normalized to $(\bar{n} \sigma v_1 2/N)^{-1}$. $v_1$ is
the magnitude of
velocity of a particle in a first excited state as defined in the
text.\label{fiSQK5}}
\end{center}
\end{figure}

\begin{figure}
\begin{center}
\caption{Fluctuation of the condensate fraction versus temperature in
thermal 
equilibrium for the 3D square well potential. Results of the
simulation 
for $N=500$. The crosses give the results from the numerical
summation of Eq.~(\protect\ref{flu}).
The dashed line is $\protect\sqrt{N_c}$ which would be equal to
$\sigma(N_c)$ if the fluctuations in the
condensate were Poissonian. \label{fiSQK2}}
\end{center}
\end{figure}

\begin{figure}
\begin{center}
\caption{Condensate fraction versus temperature in thermal
equilibrium for
the 3D harmonic oscillator.
(a) thermodynamic limit, (b) grand canonical solution for $N=500$
(solid line)
and results from the simulation (+), (c) grand canonical solution for
$N=300$ (solid line) and results from the simulation
(o).\label{fiSQK6}}
\end{center}
\end{figure}

\begin{figure}
\begin{center}
\caption{Mean collision time per particle versus temperature for
$N=500$ for 
the harmonic oscillator. Result of the simulation (solid line) and
result for 
$t_{\rm coll}=(\sigma \bar{n}^h v_{T})^{-1}$ (dashed line). The
dotted
line shows the collision time with the assumption of a fixed density
equal to the ground state density.
The time scale is normalized to $(\bar{n}^0_h \sigma v_0 2 /
N)^{-1}$.
$v_0$ is the amount of velocity of a particle in the ground state as
defined 
in the text.\label{fiSQK8}}
\end{center}
\end{figure}

\begin{figure}
\begin{center}
\caption{Mean collision time $t_{\rm{coll}}$ versus time $t$ for
$\gamma=1/10$ 
(solid line) $\gamma=1/2$ (dashed line) and $\gamma=3/2$ 
(dotted line) against time $t$.
$\gamma$ is the time constant from Eq.~(\protect\ref{board})
normalized to 
$\sigma \bar{n}_h^0 v_0 2 / N$. The time $t$ is normalized to 
$(\sigma \bar{n}_h^0 v_0 2/ N)^{-1}$.\label{fiSQKE}}
\end{center}
\end{figure}

\begin{thebibliography}{10}
\bibitem{Tmatrix}
The description of the system can be extended from a binary collision
model to
  a T-matrix description of the interaction between the particles.
\bibitem{senseBEC}
Note that we concentrate a whole energy region into the one-particle
ground
  state of the system. Therefore, onset of BEC means to get a
macroscopic
  occupation in this energy region not in a single quantum state.
\end{thebibliography}
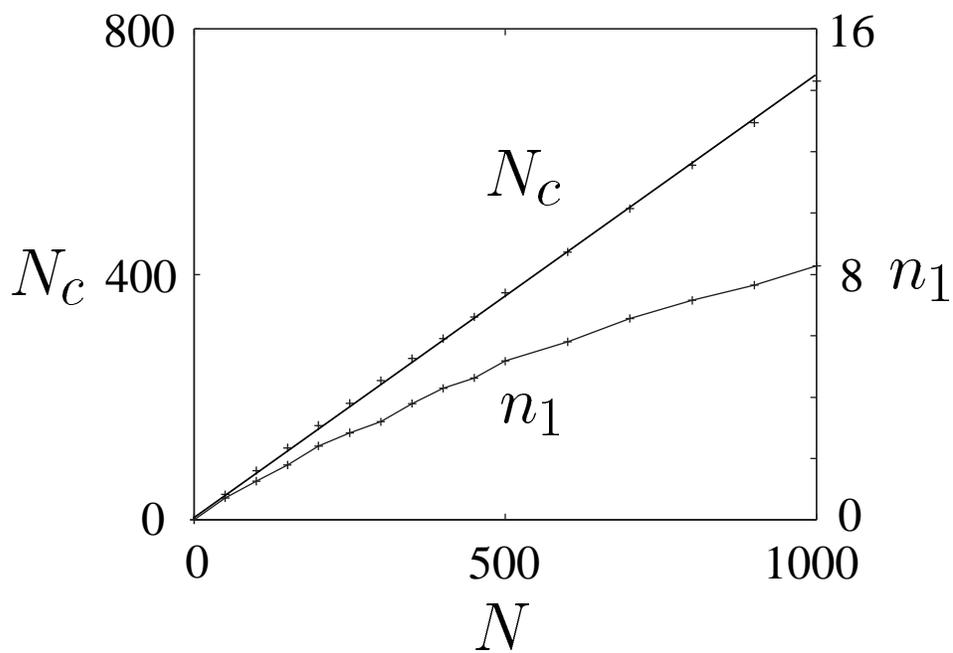
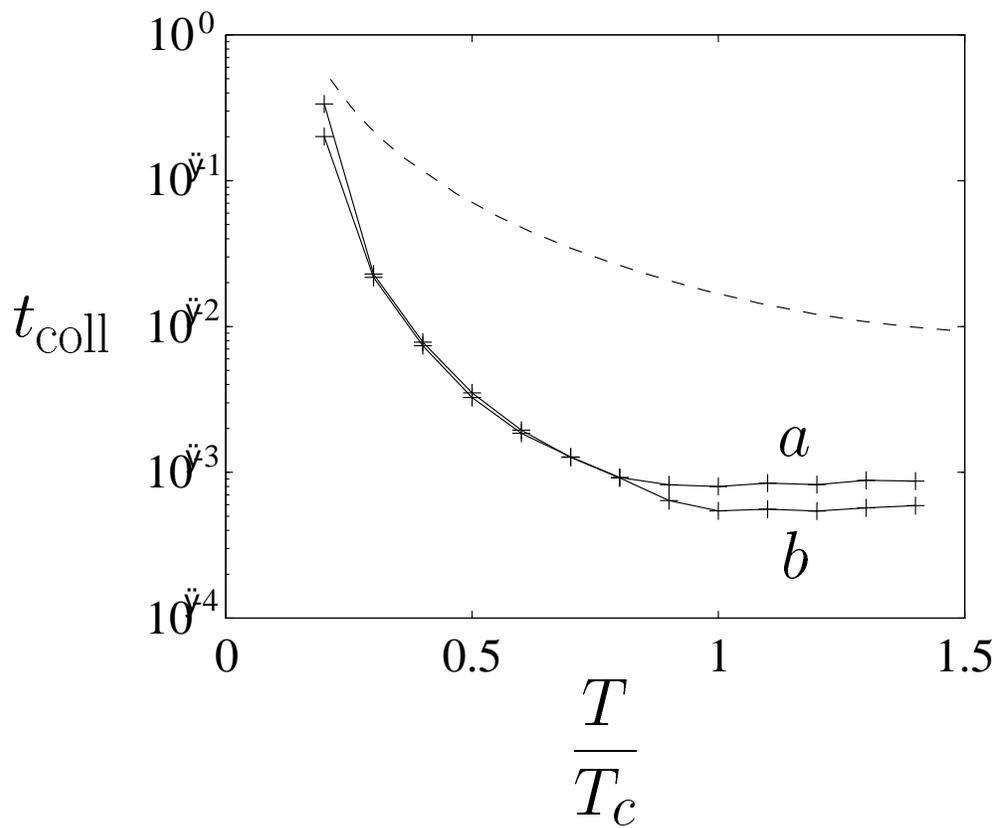
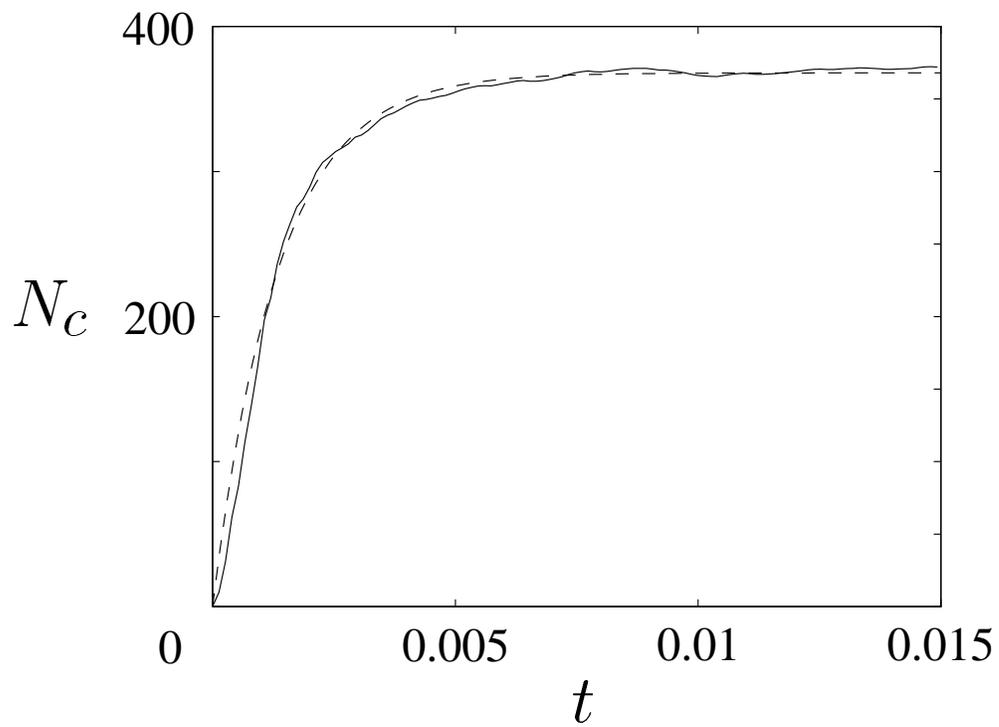
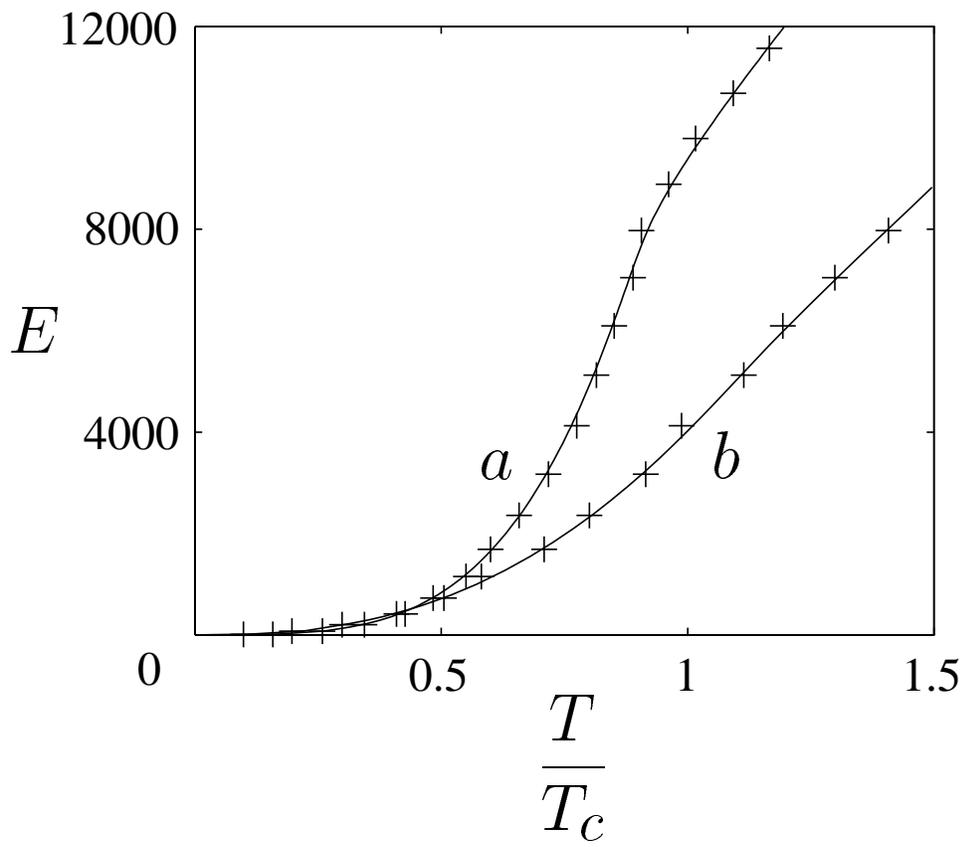
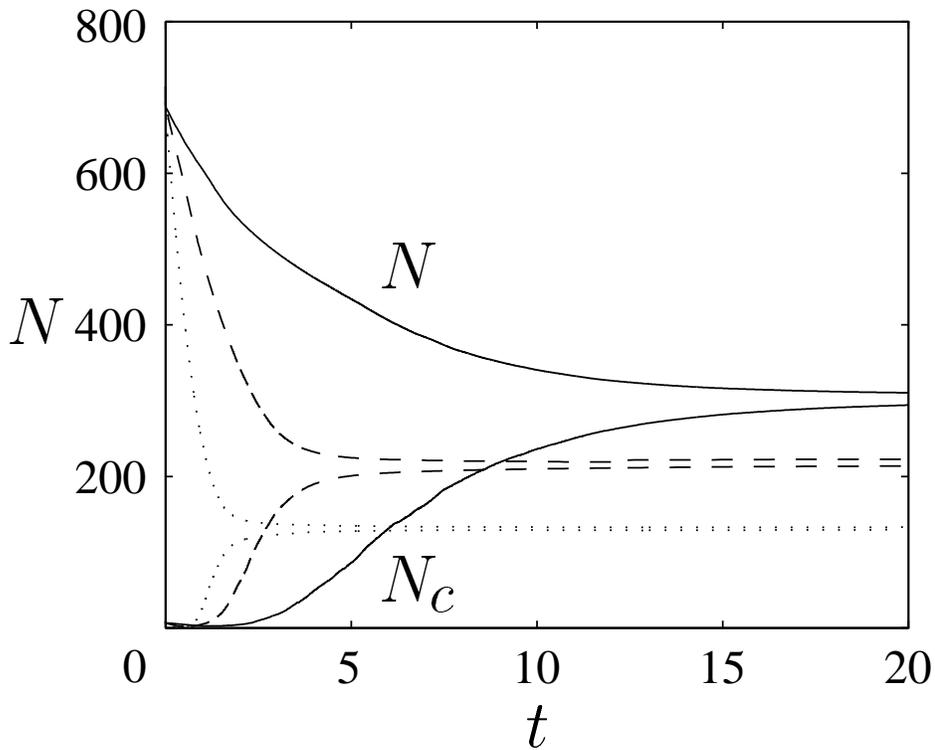
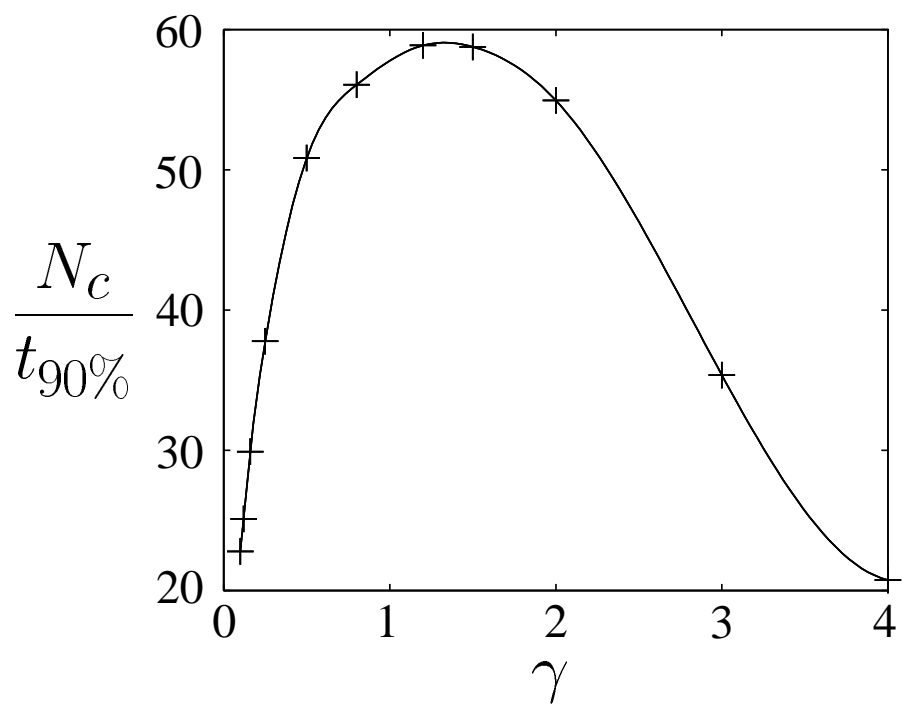
\end{document}